\documentclass[a4paper,10pt]{article}
\usepackage{times}
\usepackage{amsmath,amssymb}
\usepackage[textwidth=12.2cm,textheight=19.3cm,papersize={15.2cm,22.3cm}]{geometry}

\usepackage[square,numbers,sort&compress]{natbib}
\usepackage{bm}
\usepackage{graphicx}
\usepackage{hyperref}
\hypersetup{pdfstartview={FitH},pdfpagemode={UseNone},
            colorlinks,linkcolor=blue, citecolor=blue, urlcolor=blue,
            bookmarksopen=true, pdfnewwindow=true}

\newcommand{\eps}{\varepsilon}
\newcommand{\rmi}{{\rm i}}
\newcommand{\rmd}{{\rm d}}
\newcommand{\rot}{\mathop{\mathrm{rot}}\nolimits}
\newcommand{\e}{{\rm e}}

\newcommand{\sign}{\mathop{\mathrm{sign}}\nolimits}

\renewcommand{\Im}{\mathop{\mathrm{Im}}\nolimits}
\renewcommand{\Re}{\mathop{\mathrm{Re}}\nolimits}

\newcommand\hrefBib[3][]{\href{#2}{#3}} 

\title{Quantum nonlinear metasurfaces}
\author{\parbox{\linewidth}{
\centering
Alexander~N.~Poddubny$^{1,2,3}$,\\
Dragomir~N.~Neshev$^{4,5}$, Andrey~A.~Sukhorukov$^{1,5}$\\~\\
$^{1}$Nonlinear Physics Centre, Research School of Physics,\\The Australian National University, Canberra, ACT 2601, Australia\\
$^{2}$ITMO University, 49 Kronverksky Pr., Saint Petersburg 197101, Russia\\
$^{3}$Ioffe  Institute, Saint Petersburg 194021, Russia\\
$^{4}$Department of Electronic Materials Engineering, Research School of Physics, The Australian National University, Canberra, ACT 2601, Australia\\
$^{5}$ARC Centre of Excellence for Transformative Meta-Optical Systems (TMOS)
}
}

\begin{document}
\sloppy
\maketitle
\vspace*{-30pt}
\tableofcontents

\newpage
\section{Introduction}\label{sec:intro}

The recent advent of optical metasurfaces, ultra-thin structures composed of nanoresonators facilitating strong light concentration~\cite{Genevet:2017-139:OPT, Kuznetsov:2016-846:SCI, Neshev:2018-58:LSA}, opens the path towards efficient frequency conversion~\cite{Shcherbakov:2014-6488:NANL, Yang:2015-7388:NANL, Grinblat:2017-953:ACSN, Carletti:2015-26544:OE} in a material thousand times thinner than a human hair~\cite{Camacho-Morales:2016-7191:NANL}. There is a remarkable capacity to individually shape each nanoresonator of the metasurface to spatially and spectrally control the conversion process with unprecedented nanoscale resolution, facilitating an ultimate flexibility to selectively convert, focus, and
image different colours with a single metasurface. 
The strong enhancement of the nonlinear processes in dielectric nanoresonators is largely due to the absence of material absorption and the excitation of Mie-type bulk resonances~\cite{Kuznetsov:2016-846:SCI}. The highest conversion efficiency to date has been achieved employing III-V semiconductor nanostructures, such as AlGaAs which is a non-centrosymmetric material with high quadratic nonlinear susceptibility. In particular, second-harmonic generation efficiencies up to $10^{-4}$ have been recently demonstrated~\cite{Gili:2016-15965:OE, Liu:2016-5426:NANL, Camacho-Morales:2016-7191:NANL, Carletti:2017-114005:NANT, Liu:2018-2507:NCOM}, six orders of magnitude higher than in plasmonics.
Such capabilities can make an immense fundamental and practical impact, including the quantum state generation in nonlinear metasurfaces as we discuss in this chapter.

\begin{figure}[b!]
\begin{center}
\includegraphics[width=0.6\textwidth]{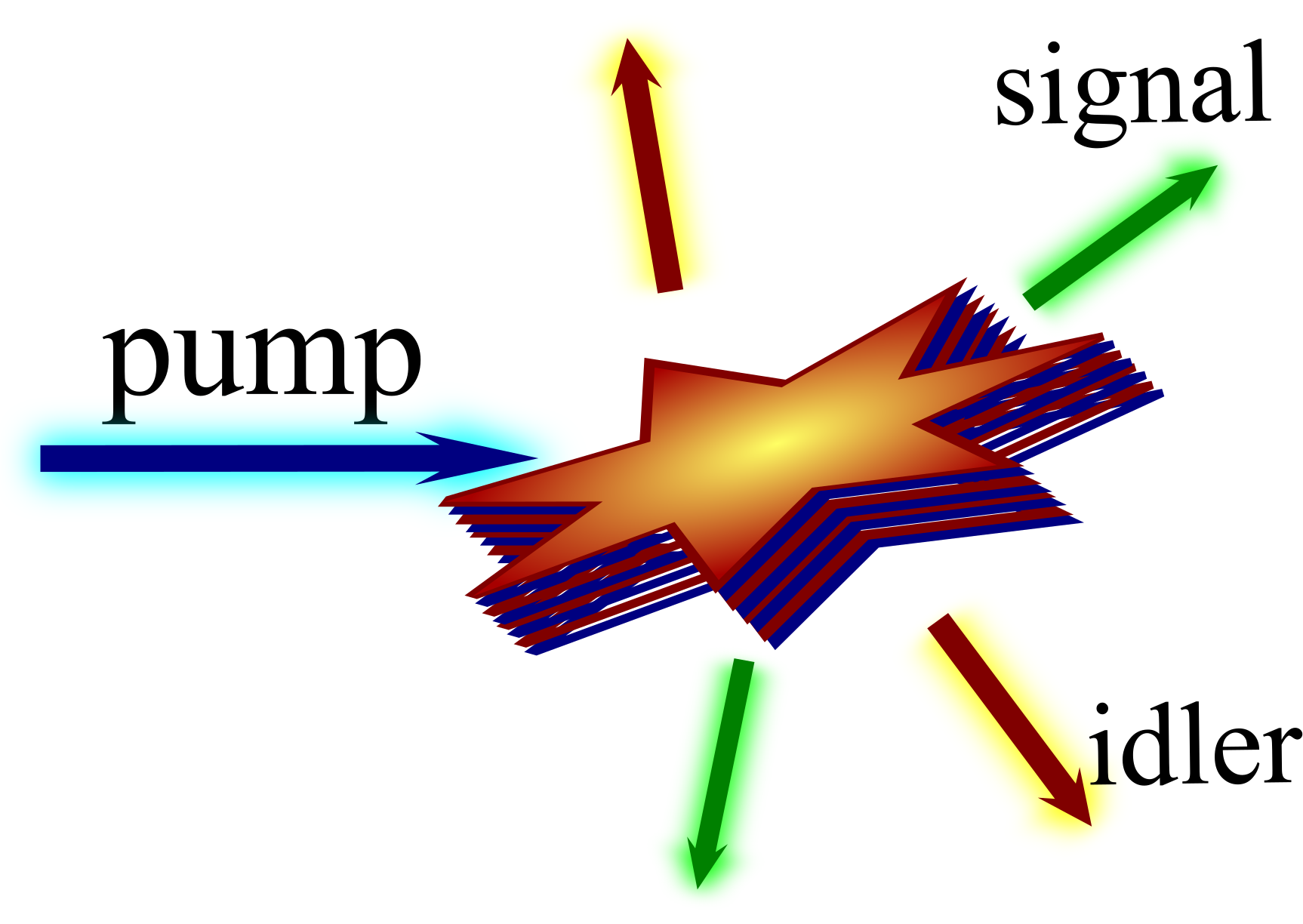}
\end{center}
\caption{General concept of bi-photon generation in an SPDC process. After Ref.~\cite{Poddubny:2016-123901:PRL}.}\label{fig:concept}
\end{figure}

The quantum state of correlated photon-pairs is the essential building block for photon entanglement~\cite{Muller:2014-224:NPHOT,Versteegh:2014-5298:NCOM}, which underpins many quantum applications, including secure networks, enhanced measurement and lithography, and quantum information processing~\cite{OBrien:2009-687:NPHOT}.
One of the most versatile techniques for the generation of correlated photons is the process of spontaneous parametric down-conversion (SPDC)~\cite{Klyshko:1988:PhotonsNonlinear, Kwiat:1995-4337:PRL}, see Fig.~\ref{fig:concept}. 
It allows for an arbitrary choice of energy and momentum correlations between the generated photons, robust operation at room temperature, as well as for spatial and temporal coherence between simultaneously pumped multiple SPDC sources. 
Alternative approaches based on atom-like single photon emitters, such as solid-state fluorescent atomic defects~\cite{Sipahigil:2016-847:SCI}, quantum dots~\cite{Senellart:2017-1026:NNANO, Somaschi:2016-340:NPHOT}, and 2D host materials~\cite{Aharonovich:2016-631:NPHOT, Tran:2016-37:NNANO}, have reached a high degree of frequency indistinguishability, purity and brightness~\cite{Somaschi:2016-340:NPHOT, Senellart:2017-1026:NNANO}. However, this comes with the expense of operation at cryogenic temperatures and lack of spatial coherence between multiple quantum emitters. These features might limit possible applications and reduce the potential for device scalability. Furthermore, the small size of the atomic sources often requires complex schemes aimed at coupling to optical nanoantennas and improving the photon extraction efficiency~\cite{Aharonovich:2016-631:NPHOT}. 

The miniaturization of SPDC quantum-light sources to micro and nanoscale dimensions is a continuing quest, as it enables denser integration of functional quantum devices. Traditionally, bulky cm-sized crystals were utilized for SPDC, entailing the difficulty of aligning multiple optical elements after the SPDC crystal, while offering relatively low photon-pair rates~\cite{Kwiat:1995-4337:PRL}. As a first step of miniaturization, SPDC was realized in low-index-contrast waveguides, which allowed confining light down to several square micrometers transversely to the propagation direction, significantly enhancing the conversion efficiency~\cite{Fiorentino:2007-7479:OE}. However, this approach still requires centimetres of propagation length, which makes the on-chip integration with other elements challenging~\cite{Solntsev:2017-19:RPH}. The introduction of high-index contrast waveguides and ring resonators allowed for shrinking the sizes necessary for SPDC to millimetres~\cite{Orieux:2013-160502:PRL}, and to tens of micrometers~\cite{Guo:2017-e16249:LSA}. However, further miniaturisation down to the metasurfaces composed of nanoresonators requires conceptually different approaches.

The generation of quantum light with nonlinear nanoresonators, acting both as sources of quantum states and nanoantennas shaping the emitted photons, has only been reported last year~\cite{Marino:2019-1416:OPT}. Such nanoscale multi-photon quantum sources offer an unexplored avenue for applications of highly indistinguishable and spatially reconfigurable quantum states, through the spatial multiplexing of coupled nanoantennas on metasurfaces. 

The chapter is organised as follows. First, in Sec.~\ref{sec:theory} we outline a general quantum theory of spontaneous photon-pair generation in arbitrary nonlinear photonic structures, including nanoresonators and metasurfaces, which provides an explicit analytical solution for the photon state expressed through the classical Green function. In the following Sec.~\ref{sec:quantum-classical} we formulate the correspondence between the quantum photon-pair generation and classical sum-frequency process in nonlinear media, and discuss its application in various contexts, including waveguide circuits and nanostructures. Then, in Sec.~\ref{sec:experiment} we present the first experimental results demonstrating photon-pair generation in a single nonlinear nanoantenna. Finally, in Sec.~\ref{sec:outlook} we present conclusions and outlook towards the generation of quantum entangled images with nonlinear metasurfaces and emerging opportunities for applications.
\section{Green function theory}\label{sec:theory}
In this section we outline the general theoretical approach to for two-photon generation via 
spontaneous parametric down conversion (SPDC) and spontaneous four-wave mixing (SFWM). Sec.~\ref{sec:GeneralGreen} presents the derivation of the two-photon wavefunction.  In Sec.~\ref{sec:herald} we discuss the photon heralding efficiency. In Sec.~\ref{sec:correspond} we show, how our results can be reduced to
those known in literature for the specific case of coupled dispersionless waveguides. Finally, in Sec.~\ref{sec:plasmons} we consider the generation of entangled plasmon-photon pairs.
\subsection{Two-photon wavefunction }\label{sec:GeneralGreen}
The photon-pair generation is described by the Hamiltonian~\cite{Drummond:2014:QuantumTheory}
\begin{equation}\label{seq:HNL}
 H_{\rm NL}=\frac{1}{2}\int \frac{{\rmd \omega_{1}\rmd \omega_{2}}}{(2\pi)^{2}}
\int\rmd^{3}rE_{\alpha}^{\dag}(\omega_{1},\bm r)E_{\beta}^{\dag}(\omega_{2},\bm r)
\Gamma_{\alpha\beta}(\bm r)+\rm H.c.\:
\end{equation}
where $E$ is the electric field operator, $\alpha,\beta=x,y,z$, and $\Gamma_{\alpha\beta}$ is the generation matrix.
We consider two possibilites~\cite{Boyd:2008:NonlinearOptics}, spontaneous parametric down conversion (SPDC) due to $\chi^{(2)}$ nonlinear susceptibility and spontaneous four-wave mixing (SFWM) governed by $\chi^{(3)}$ nonlinearity, when
\begin{equation}\label{seq:Gamma}
\Gamma_{\alpha\beta}(\bm r)=
\begin{cases}
\chi^{(2)}_{\alpha\beta\gamma}(\bm r; \omega_{1},\omega_{2};\omega_{p})\mathcal E_{p,\gamma}(\bm r) \e^{-\rmi\omega_{\rm p}t} ,\\*[9pt]
\chi^{(3)}_{\alpha\beta\gamma\delta}(\bm r; \omega_{1},\omega_{2};\omega_{p},\omega_{p})\mathcal E_{p,\gamma}(\bm r)\mathcal E_{p,\delta}(\bm r)\e^{-2\rmi\omega_{\rm p}t} ,\:
\end{cases}
\end{equation}
\begin{figure}[b!]
\begin{center}
\includegraphics[width=0.99\textwidth]{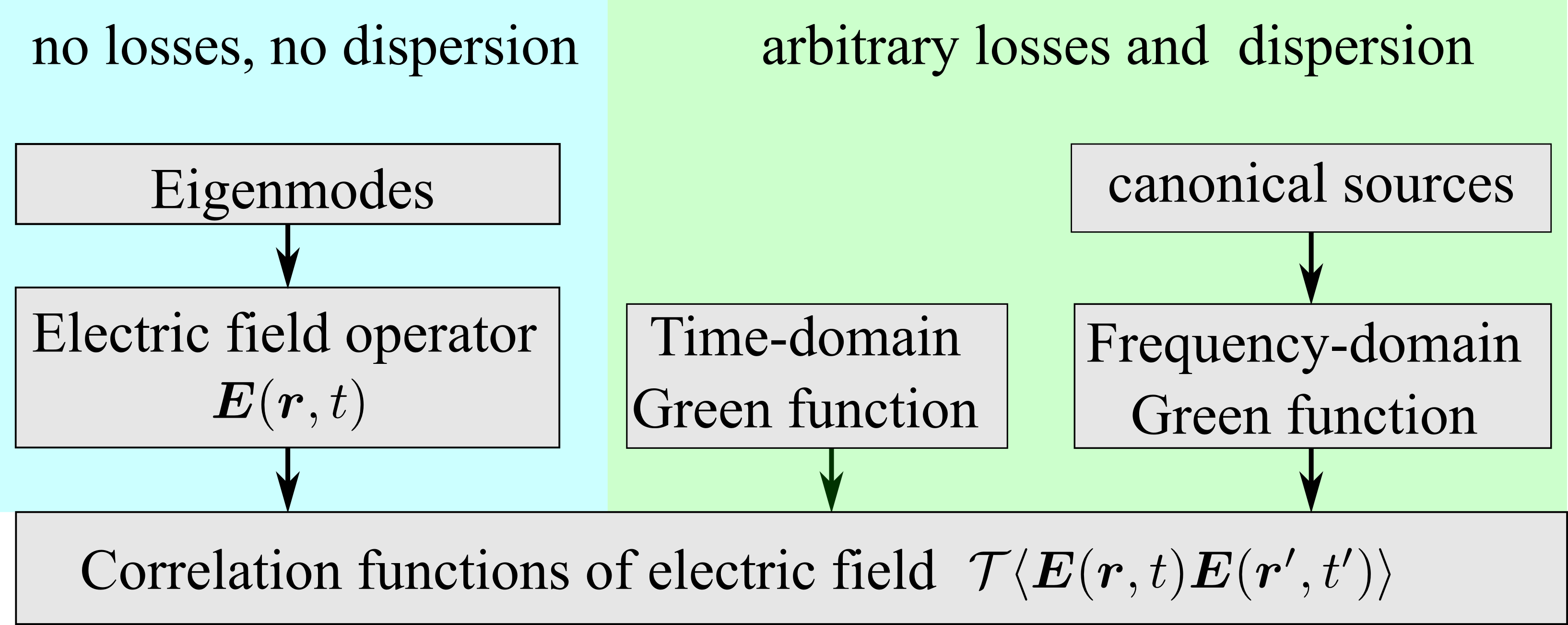}
\end{center}
\caption{Scheme of different approaches to calculate the two-photon wavefunction}\label{fig:CalcScheme}
\end{figure}

Here, $\mathcal E_{p}$ is the classical pump at frequency $\omega_{\rm p}$, and $\gamma,\delta=x,y,z$ are the Cartesian indices. 

We explicitly introduce the {\em sensors that detect the quantum electromagnetic field}~\cite{delValle:2012-183601:PRL} to find the experimentally measurable quantities.
The sensors are modelled as  signal (s) and idler (i) two-level systems with the Hamiltonians
\begin{equation}
H_{i,s}\equiv H_{i,s}^{(0)}+V_{i,s}=\hbar\omega_{i,s}a_{i,s}^{\dag}a_{i,s}-\hat{\bm d}_{i,s}\cdot \bm E(\bm r_{i,s})\:,
\end{equation}
with the resonant energies $\hbar\omega_{s}$ and $\hbar\omega_{i}$, respectively. Here,  $a^{\dag}_{s,i}$ are the corresponding exciton creation and
$\hat{\bm d}_{i,s}=a\bm d^{\ast}_{i,s}+a^{\dag}\bm d_{i,s}^{\ast}$ are the dipole momentum operators.
The detected two-quantum state is $|\Psi\rangle=a_{i}^{\dag}a_{s}^{\dag}|0\rangle$ with both detectors excited by the photon pair.

The direct approach to describe the two-photon generation would be to expand the electric field operator over the set of eigenmodes of the linear problem, see the left part Fig.~\ref{fig:CalcScheme}. However, this technique is impractical for nanostructured metamaterials. The eigenmodes are poorly  defined due to the presence of Ohmic and radiative losses as well as the dispersion. Instead,  we describe the linear electromagnetic problem by the classical Green tensor,
\begin{equation}
\label{seq:G}
 [\rot\rot-\left(\frac{\omega}{c}\right)^2\varepsilon(\omega,\bm r)]G(\bm r,\bm r',\omega)=4\pi \left(\frac{\omega}{c}\right)^2 \hat 1\delta(\bm r-\bm r')\:,
\end{equation}
that explicitly accounts for arbitrary strong Ohmic losses and mode dispersion. The {\it classical} Green function 
allows to circumvent the calculation of eigemodes in macroscopic {\it quantum} electrodynamics. 
The Green function method was previously applied~\cite{Poddubny:2012-33826:PRA} to describe the spontaneous two-photon emission (STPE)~\cite{Hayat:2008-238:NPHOT} from a single atom. However the current problem is quite distinct from STPE, because nonlinear spontaneous wave mixing acts as a coherent spatially extended source.

In the  Green function method the  two-photon wavefunction  $|\Psi\rangle$ can be equivalently obtained using the time-dependent perturbation theory or the time-independent one, see middle and right parts in Fig.~\ref{fig:CalcScheme}, respectively. The first approach  follows the methodology of the quantum field  and condensed matter theories and allows for a more compact derivation provided that the  Feynman diagram technique~\cite{landau9}  is used. The second one requires only  the knowledge of standard quantum mechanics but has to be supplemented with  the  local source quantization scheme for the electric field~\cite{Vogel:2006:QuantumOptics}.
 Below we will outline the derivation using both methods and demonstrate their equivalence.
\subsubsection{Time-dependent perturbation technique}

Formally, the process of photon pair generation, propagation, and detection can be described by the scattering matrix element $S_{is}=\langle \Psi|U|0\rangle$, where $U$ is the evolution operator~\cite{Cohen:1998:AtomPhoton},\begin{equation}
U=\mathcal T\e^{-\rmi\int_{-\infty}^{\infty} W(\tau)\rmd \tau}.
\end{equation}
Here, $\mathcal T$ denotes the time-ordered product and
\begin{equation}
W(\tau)=\e^{\rmi H^{(0)}\tau/\hbar}(V_i+V_s+ H_{\rm NL})\e^{-\rmi H^{(0)}\tau/\hbar},\quad H^{(0)}=H^{(0)}_i+H^{(0)}_s+H_{\rm lin}\: \label{seq:V}
\end{equation}
is the operator describing the generation and detection in the interaction representation and $H^{(0)}$ is the sum of  
Hamiltonian of noninteracting photons $H^{(0)}_{i}$ and the detector Hamiltonians. Our goal is to calculate the scattering matrix element in the lowest nonvanishing order of the time-dependent perturbation theory.
Since two photons have to be  generated and each of them has to be absorbed, we need to consider the third order process in the operator Eq.~\eqref{seq:V}, given by the expansion
\begin{equation}
S_{is}=\left(\frac{-\rmi}{\hbar}\right)^{3}\frac1{3!}\int\limits_{-\infty}^{\infty}\int\limits_{-\infty}^{\infty}\int\limits_{-\infty}^{\infty}\rmd \tau_{1}\rmd\tau_{2}\rmd \tau_{3} \mathcal T \langle \Psi| W(\tau_{1})W(\tau_{2})W (\tau_{3})|0\rangle\:.\label{seq:216}
\end{equation}
\begin{figure}[b!]
\begin{center}
\includegraphics[width=0.99\textwidth]{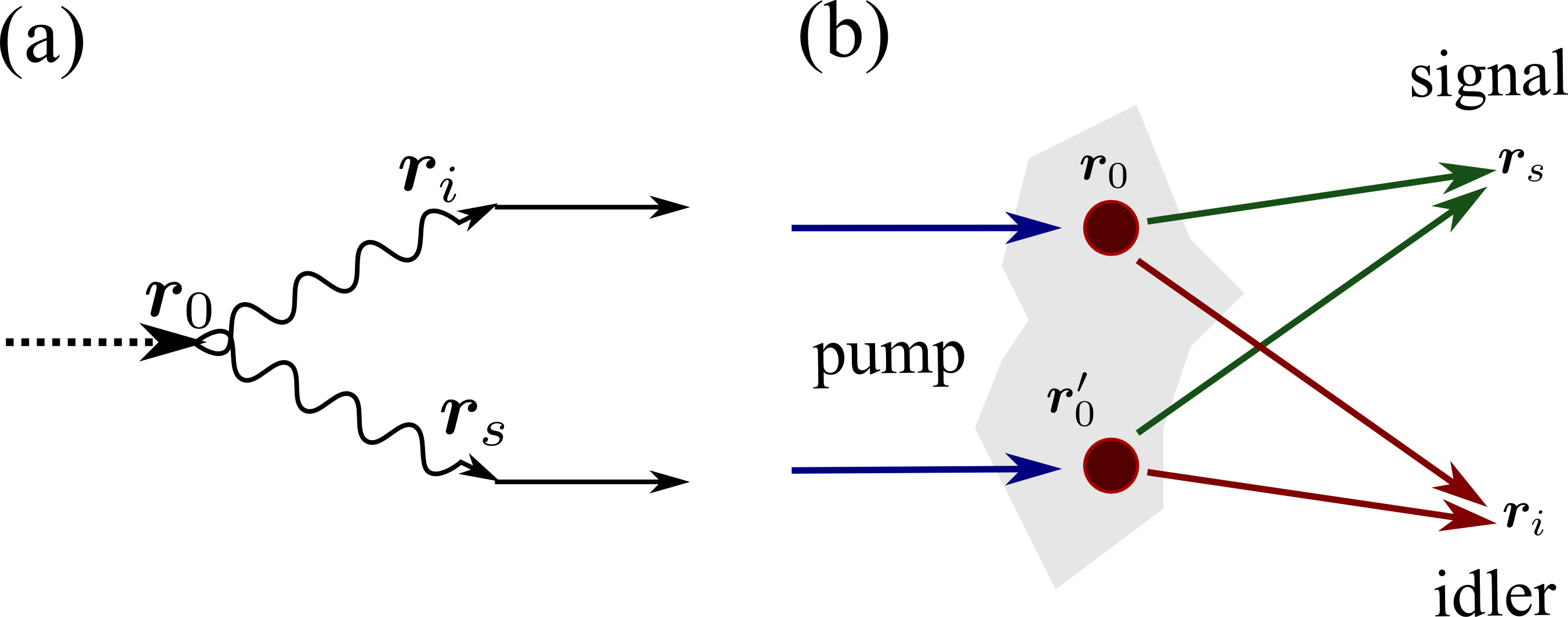}
\end{center}
\caption{(a)~Feynman diagram for Eq.~\eqref{seq:216}, describing generation of two photons in the point $\bm r_0$ and their detection in the points $\bm r_i$ and $\bm r_f$. Dotted line corresponds to the classical pump, curved lines describe photon Green functions and solid line are the  detectors. (b)~Graphic representation of the same process and Eq.~\eqref{seq:Tis113} for the two-photon wave function. After Ref.~\cite{Poddubny:2016-123901:PRL}.}\label{fig:S1}
\end{figure}

\noindent The diagrammatic representation of Eq.~\eqref{seq:216} is shown in Fig.~\ref{fig:S1}(a).
Explicitly, the matrix element reads
\begin{multline}
\mathcal T \langle\Psi| W(\tau_{1})W(\tau_{2})W (\tau_{3})|0\rangle\\=\frac{1}{2}d^\ast_{i}d^\ast_{s}
\e^{\rmi \tau_1 \omega_i+\rmi \tau_2 \omega_s-N\rmi\omega_{\rm p}\tau_3}\\
\times \sum\limits_{\alpha\beta}
\Gamma_{\alpha\beta}(\bm r_0)\int \rmd^3 r_0
 \mathcal T\langle E_{\sigma_i}(\bm r_{i},\tau_{1})E_{\sigma_s}(\bm r_{s},\tau_{s})E_\alpha(\bm r_{0},\tau_{3})E_\beta(\bm r_{0},\tau_{3})\rangle\\+\text{permutations of $\tau_1$,$\tau_2$,$\tau_3$,}\label{seq:T2}
\end{multline}
where $N=1 (2)$ for SPDC (SWFM).
As usual, the factor $1/3!$ in Eq.~\eqref{seq:216} cancels out with the combinatorial factor arising from 6 different permutations of the times $\tau_1,\tau_2,\tau_3$~\cite{landau9}.
Hence, the scattering matrix element is determined only by the correlation function in the second line of Eq.~\eqref{seq:T2}.

The correlation functions  in Eq.~\eqref{seq:T2} are calculated by using  the equivalent alternative definition of the Green function from Eq.~\eqref{seq:G} in the time domain~\cite{landau9}
\begin{equation}\label{seq:G1}
G_{\alpha\beta}(\bm r,\bm r',t-t')=\frac\rmi{\hbar} \mathcal T\langle E_{\alpha}(t,\bm r)E_{\beta}(t',\bm r')\rangle \:.
\end{equation}
The function in Eq.~\eqref{seq:G1} is related to the frequency-dependent function in Eq.~\eqref{seq:G} by the Fourier transform $G(t)=\int\rmd\omega\e^{-\rmi \omega t}G(\omega)/2\pi$.  Once the Green function is introduced, the correlation function  in Eq.~\eqref{seq:T2} can be calculated by means of the Wick's theorem,
\begin{multline}\label{seq:T2a}
 \mathcal T\langle E_{\sigma_i}(\bm r_{i},\tau_{1})E_{\sigma_f}(\bm r_{s},\tau_{2})E_\alpha(\bm r_{0},\tau_{3})E_\beta(\bm r_{0},\tau_{3})\rangle\\=
\mathcal T\langle E_{\sigma_i}(\bm r_{i},\tau_{1})E_{\alpha}(\bm r_{0},\tau_{3})\rangle\langle E_{\sigma_s}(\bm r_{s},\tau_{2})E_{\beta}(\bm r_{0},\tau_{3})\rangle+\\
\mathcal T\langle E_{\sigma_i}(\bm r_{i},\tau_{1})E_{\sigma_s}(\bm r_{s},\tau_{2})\rangle\langle E_{\alpha}(\bm r_{0},\tau_{3})E_{\beta}(\bm r_{0},\tau_{3})\rangle
]\:.
\end{multline}
Using the Green function definition in Eq.~\eqref{seq:G1} we find
 the scattering matrix element in the form
\begin{equation}\label{seq:Sis}
S_{is}=-2\pi \rmi \delta(\hbar\omega_{i}+\hbar\omega_{s}-N\hbar\omega_{\rm pump})T_{is}
\end{equation}
 and  obtain
\begin{multline}
T_{is}(\bm r_{i},\omega_{i},\bm d_{i};\bm r_{s},\omega_{s},\bm d_{s})=\sum\limits_{\alpha\beta,\sigma_{i},\sigma_{s}}d^{\ast}_{i,\sigma_{i}}d^{\ast}_{s,\sigma_{s}}\\\times\int \rmd^{3} r_{0}
G_{\sigma_{i}\alpha}(\bm r_{i},\bm r_{0},\omega_{i})
G_{\sigma_{s}\beta}(\bm r_{s},\bm r_{0},\omega_{s})
\Gamma_{\alpha\beta}(\bm r_{0})\:.\label{seq:Tis113}
\end{multline}
By construction the  two-photon transition amplitude  $T_{is}$ has the meaning of the complex wave function fully defining the pure two-photon state. 

Equation \eqref{seq:Tis113} is the central result of our study. The form of Eq.~\eqref{seq:Tis113} clearly represents the interference between the spatially entangled photons generated in the different points of space $\bm r_{0}$~\cite{Ghosh:1987-1903:PRL}, as schematically illustrated in Fig.~\ref{fig:S1}(b).
The coincidence rate, which defines simultaneous detection of two photons at different positions in space, is found as:
\begin{equation}
W_{is}=(2\pi/\hbar)\delta(\hbar \omega_{i}+\hbar\omega_{s}-N\hbar\omega_{p})|T_{is}|^{2}\:. 
\end{equation}
\subsubsection{Time-independent local source quantization scheme}
In this approach instead of Eq.~\eqref{seq:216} in the time domain we write \cite{Cohen:1998:AtomPhoton}
for the two-photon wavefunction  
\begin{equation}
T_{is} =\lim_{\eps\to 0}\langle \Psi|V\frac{1}{N\hbar\omega_{p}-H_{0}+\rmi \eps}V\frac{1}{N\hbar\omega_{p}-H_{0}+\rmi \eps}V|0\rangle\:,\label{eq:Tis2}
\end{equation}
in the frequency domain, where $N=1,2$ for SPDC (SFWM), respectively.
The Hamiltonian $H_{\rm lin}$, describing the  {\em linear propagation of the generated photons}, is written 
in the {\it local source quantization scheme}~\cite{Vogel:2006:QuantumOptics}  as
\begin{equation}
H_{\rm lin}=\int \rmd^{3}r\int\limits_{0}^{\infty}\sum_{\alpha=x,y,z}\rmd \omega\hbar \omega f_\alpha^{\dag}(\bm r,\omega) f_\alpha^{\vphantom{\dag}}(\bm r,\omega)\:,
\end{equation}
where $f_{\alpha}(\bm r,\omega)$ are the canonical bosonic source operators for the quantum electric field~\cite{Vogel:2006:QuantumOptics}:
\begin{align}\label{seq:E}
&\bm E(\bm r)=\int\limits_{0}^{\infty}\frac{\rmd \omega}{2\pi}\bm E(\bm r,\omega)+{\rm H.c.}\:,\\
&\hat{\bm E}(\omega)= \rmi\sqrt{\hbar} \int \rmd^3\bm r'  G_{\alpha\beta}(\bm r,\bm r',\omega)\sqrt{\Im\eps(\omega,\bm r')}f_{\beta}(\bm r',\omega).\nonumber
\end{align}
Now we directly substitute Eq.~\eqref{seq:E} into Eq.~\eqref{eq:Tis2}. The averaging  yields the two products of the Green functions, that are evaluated using the identity \cite{Vogel:2006:QuantumOptics}
\begin{equation}\label{eq:identity}
\Im G_{\beta\beta'}(\bm r,\bm r'')=\frac{1}{4\pi}\int \rmd^3r' \Im[\eps(\bm r')]G_{\beta\alpha}(\bm r,\bm r')
G^{\ast}_{\beta'\alpha'}(\bm r'',\bm r')\:,
  \end{equation}
and the integration over frequency $\omega'$ is performed  using the Kramers-Kronig relations,
  \begin{equation}
\lim_{\eps\to 0}  \int_{0}^{\infty}\frac{d\omega'}{\pi}\Im G_{\alpha\beta}(\bm r,\bm r')\left(\frac1{\omega'-\omega_{s}-\rmi \eps}+
\frac1{\omega_{s}+\omega'-\rmi \eps}\right)=G_{\alpha\beta}(\bm r,\bm r')\:.\label{eq:KK}
  \end{equation}
 The final result yields again Eq.~\eqref{seq:Tis113} in full agreement with the time-dependent technique.
 The time-dependent approach relies directly on the Green function from Eq.~\eqref{seq:G1} while the time-independent one is built on  the electric field expansion in Eq.~\eqref{seq:E}. Their equivalence can be verified by substituting 
 Eq.~\eqref{seq:E} into Eq.~\eqref{seq:G1}.
\subsection{Heralding efficiency}\label{sec:herald}
The role of ohmic and radiative losses becomes especially important for the two-photon generation process. After an entangled photon pair  is generated, a single absorption event is sufficient to fully destroy it, and leave only one photon in a mixed state. Hence, the problem of detection of individual photons in a lossy system where entangled photon pairs are generated, provides an interesting question. Given that a signal photon is detected, did  it correspond to the photon pair where the idler photon has been emitted or to the pair, where it has been absorbed? The relative magnitude of the first term provides the  heralding efficiency, characterizing the robustness of the two-photon generation setup to the ohmic losses.

In order to evaluate the heralding efficiency  we start with the calculation of the total single-photon count rate. 
Similarly to the two-photon wavefunction, the rate can be rigorously obtained using the time-dependent perturbation theory in the framework of the Keldysh diagram technique  or using time-independent approach with the canonical local sources~\cite{Poddubny:2016-123901:PRL}. Another equivalent and very compact derivation, given below, is based on the  semiclassical expression for the signal polarization combined with the fluctuation-dissipation theorem.
We use the standard result of the quantum  photodetector theory  for the signal photon count rate \cite{ScullyZubairy,Carmichael}
\begin{equation}
 W_s(\bm r_s,\sigma_s)=\frac{1}{\hbar}\sum\limits_{\sigma_s\sigma_{s'}} d^{\vphantom{\ast}}_{s\sigma_s}d^\ast_{s',\sigma_s'}\int\limits_{-\infty}^{\infty}\langle E_{\sigma_s}(\bm r_{s},\tau)E_{\sigma_{s'}}(\bm r_{s},0)\rangle\e^{\rmi\omega_{s}\tau}\rmd \tau\:, \label{seq:Ws}
\end{equation}
where the averaging is performed over the states of the electromagnetic field $E$. In order to calculate the field of signal photons, acting upon the detector, we  introduce the dielectric polarization  as a variation of the density of the nonlinear Hamiltonian in Eq.~\eqref{seq:HNL} over the electric field
\begin{equation}
P_{\alpha}(\bm r_0,t)=-\frac{\delta H_{\rm NL}}{\delta E_s}=
-\Gamma_{\alpha\beta}(\bm r_0)E_{i,\beta}(\bm r_0,t)\:.\label{seq:Ps}
\end{equation}
Next, we determine the electric field induced by this polarization
\begin{equation}
E_{\sigma_s}(\bm r_s,t)=\int\limits_{-\infty}^t\rmd t' G_{\sigma_s\alpha}(\bm r_s,\bm r_0,t-t')P_{\alpha}(\bm r_0,t')\:,\label{seq:Es}
\end{equation}
where $G$ is the time-dependent retarded Green function.
 Using Eq.~\eqref{seq:Ps}, Eq.~\eqref{seq:Es}  and the fluctuation-dissipation theorem \cite{landau5} for the electric field components in the form~\cite{landau9}
\begin{equation}
\langle
E_{\beta}(\bm r,t)E_{\beta'}(\bm r',t')
\rangle=\int\limits_{-\infty}^{\infty} \frac{\rmd \omega}{2\pi}\hbar \Im G^R_{\beta\beta'}(\omega,\bm r_{1},\bm r_{2})\sign \omega
\e^{-\rmi\omega(t-t')}
\end{equation}
we obtain the signal photon count rate 
\begin{multline}\label{seq:Ws_final}
W_{s}(\bm r_{s})= \frac{2}{\hbar}\iint \rmd^{3} r_{0}'\rmd^{3} r_{0}''\sum\limits_{\sigma_{s\vphantom{'}}\sigma_{s'}} d_{s,\sigma_{s}}
d^\ast_{s,\sigma_{s'}}
\Im G_{\beta\beta'}(\bm r_{0}',\bm r_{0}'',\omega_{p}-\omega_{s})\\\times
\Gamma_{\alpha\beta}(\bm r_{0}')\Gamma^{\ast}_{\alpha'\beta'}(\bm r_{0}'')G_{\sigma_{s},\alpha}(\bm r_{s},\bm r_{0}',\omega_{s})G_{\sigma_{s'},\alpha'}^{\ast}(\bm r_{s},\bm r_{0}'',\omega_{s})\:.
\end{multline}
\begin{figure}[t]
\begin{center}
\includegraphics[width=0.99\textwidth]{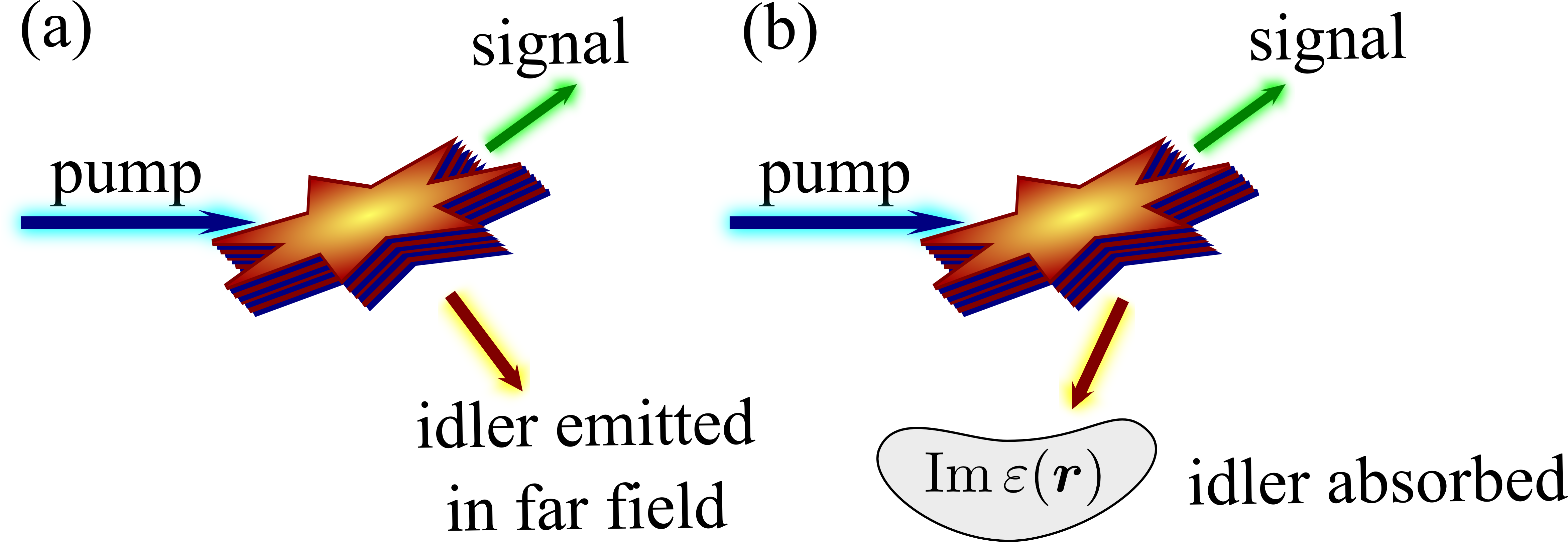}
\end{center}
\caption{Illustration of the photon pairs with idler photon (a)~emitted and (b)~absorbed. The relative value of the pairs (a), corresponding to the first term  in  Eq.~\eqref{eq:Ws12}, provides the heralding efficiency.}\label{fig:farnear}
\end{figure}

\noindent It is instructive to separate the rate of single photon counts in  Eq.~\eqref{seq:Ws_final} into two contributions,
\begin{equation}
W_s(\bm r_s)=W_s^{(\rm rad)}(\bm r_s)+W_s^{(\rm Ohmic)}(\bm r_s)\:.\label{eq:Ws12}
\end{equation}
The first term corresponds to the photon pairs with  the idler photon emitted into the far field, and the second term describes the pairs with  the idler  photon absorbed in the medium, see Fig.~\ref{fig:farnear}(a) and Fig.~\ref{fig:farnear}(b), respectively. The ratio of the  first term to the total single-photon count rate will provide the far-field heralding efficiency.
Explicitly,   the radiative term reads
\begin{align}
W^{\rm (rad)}_{s}(\bm r_{s},\sigma_{s})=&\nonumber
\frac{c }{2\pi \hbar \omega_{i}}
\sum\limits_{\sigma_s,\sigma_{s'},\beta}d_{s,\sigma_s}d_{s,\sigma_{s'}}^\ast
\oint \rmd S_{i,\beta}e_{\beta \sigma_i\sigma_{i'}}\\\nonumber& \Re
\tilde{{ T}}_{is}(\bm r_{s},\bm r_{i},\sigma_s,\sigma_i)\bar{{ T}}_{is}^\ast(\bm r_{s},\bm r_{i},\sigma_{s'},\sigma_{i'})\:,\\
\tilde T_{is}(\bm r_{s},\bm r_{i},\sigma_s,\sigma_i)&\label{seq:Wrad}=\int \rmd^{3} r_{0}
G_{\sigma_{i}\alpha}(\bm r_{i},\bm r_{0},\omega_{i})
G_{\sigma_{s}\beta}(\bm r_{s},\bm r_{0},\omega_{s})
\Gamma_{\alpha\beta}(\bm r_{0})
,\\
\bar {T}_{is}(\bm r_{s},\bm r_{i},\sigma_{s},\sigma_{i})&=
\frac{c}{\rmi \omega_{i}}\sum\limits_{\beta\gamma}e_{\sigma_i\beta\gamma}\frac{\partial \tilde T(\bm r_{s},\bm r_{i},\sigma_s,\gamma)}{\partial{x_{i,\gamma}}}\:,\:\omega_i=N\omega_{\rm p}-\omega_s\:,\nonumber
\end{align}
where $e_{\beta \sigma_i\sigma_{i'}}$ is the Levi-Civita tensor.
The second term, corresponding to the pairs where the idler photon has been absorbed, has the form
\begin{equation}\label{seq:Ohmic}
W_s^{(\rm Ohmic)}(\bm r_s)=\frac{\hbar }{4\pi^{2}|d_{i}|^{2}}\sum\limits_{\sigma_{i}}\int \rmd\omega_{i}\int  \rmd \bm r_{i}W_{is}(\bm r_{i},\omega_{i},\bm d_i;\bm r_{s},\omega_{s}, \sigma_{s})\Im\eps(\bm r_{i},\omega_{i})\:,
\end{equation}
and can be expressed via the joint two-photon count rate
\begin{equation}\label{seq:double}
W_{is}(\bm r_{i},\omega_{i},\bm d_{i};\bm r_{s},\omega_{s}, \sigma_{s})=\frac{2\pi}{\hbar}\delta(\hbar \omega_{i}+\hbar \omega_{s}-N\hbar\omega_{\rm p})|T_{is}(\bm r_{i},\omega_{i},\bm d_{i};\bm r_{s},\omega_{s}, \bm e_{\sigma_s} d_{s})|^{2}\:.
\end{equation}
The volume integration in Eq.~\eqref{seq:Ohmic} is performed only over the lossy region where $\Im \eps\ne 0$. If the integral is extended over the whole volume and is regularized at $r\to\infty$ by adding infinitesimal losses to the permittivity as
\begin{equation}
\eps (\bm r_i)\to \eps (\bm r_i)+\rmi \Delta\eps\:,
\end{equation}
the result reduces to the total rate $W_s(\bm r_s)$ in Eq.~\eqref{seq:Ws_final}, rather than to $W_s^{(\rm Ohmic)}(\bm r_s)$:
\begin{equation}\label{seq:WsAll2}
W_s(\bm r_s)=\frac{\hbar }{4\pi^{2}|d_{i}|^{2}}\sum\limits_{\sigma_{i}}\int \rmd\omega_{i}
\lim_{\Delta\eps\to 0}
\int_{\rm tot}  \rmd \bm r_{i}W_{is}(\bm r_{i},\omega_{i},\bm d_i;\bm r_{s},\omega_{s}, \sigma_{s})\Im\eps(\bm r_{i},\omega_{i})\:.
\end{equation}

By definition, the heralding  efficiency is the ratio of the  heralded counts, corresponding to the pairs where  the idler photon has been emitted into the far field to the total single-photon rate.  As such, this quantity should characterize the generated photon pairs but not the specific detection setup. However, the single-photon count rates in Eq.~\eqref{seq:Ws_final}, Eq.~\eqref{seq:Wrad}, Eq.~\eqref{seq:Ohmic} and the two-photon count rate in Eq.~\eqref{seq:double} by construction depend on the efficiencies of the detectors, determined by the dipole momentum matrix elements $\bm d_{i,s}$. In order to determine the heralding efficiency, it is necessary to explicitly calibrate the detectors to the flux of the photons. The procedure is outlined in \cite{Poddubny:2016-123901:PRL} and the calibrated heralding efficiency  reads
\begin{align}\label{seq:QE}
QE=&\frac{\tilde W_s^{(\rm rad)}}{\tilde W_s}\:,\\
\tilde W_s^{(\rm rad)}=&\frac{c}{4\pi\omega_i}\oint\rmd S_{i,\alpha} e_{\alpha \sigma_i\sigma_{i'}} \Re
\tilde{{ T}}_{is}(\bm r_{s},\bm r_{i},\sigma_s,\sigma_i)\bar{{ T}}^\ast_{is}(\bm r_{s},\bm r_{i},\sigma_{s'},\sigma_{i'})\:,\nonumber\\
\tilde W_s=&\iint \rmd^{3} r_{0}'\rmd^{3} r_{0}''\Im G_{\beta\beta'}(\bm r_{0}',\bm r_{0}'',\omega_{p}-\omega_{s})
\Gamma_{\alpha\beta}(\bm r_{0}')\Gamma^{\ast}_{\alpha'\beta'}(\bm r_{0}'')\times \nonumber \\ &\qquad \qquad \qquad G_{\sigma_{s},\alpha}(\bm r_{s},\bm r_{0}',\omega_{s})G_{\sigma_{s},\alpha'}^{\ast}(\bm r_{s},\bm r_{0}'',\omega_{s})\:.\nonumber
\end{align}
 In the case of planar geometry, the Green functions in Eq.~\eqref{seq:QE} are to be replaced by their Fourier components along $x,y$ and the unit area interval by the unit wave vector interval, $1/{\rmd S}\to \rmd k_x\rmd k_y/(2\pi)^2$.  
\begin{figure}[t]
\begin{center}
\includegraphics[width=0.75\textwidth]{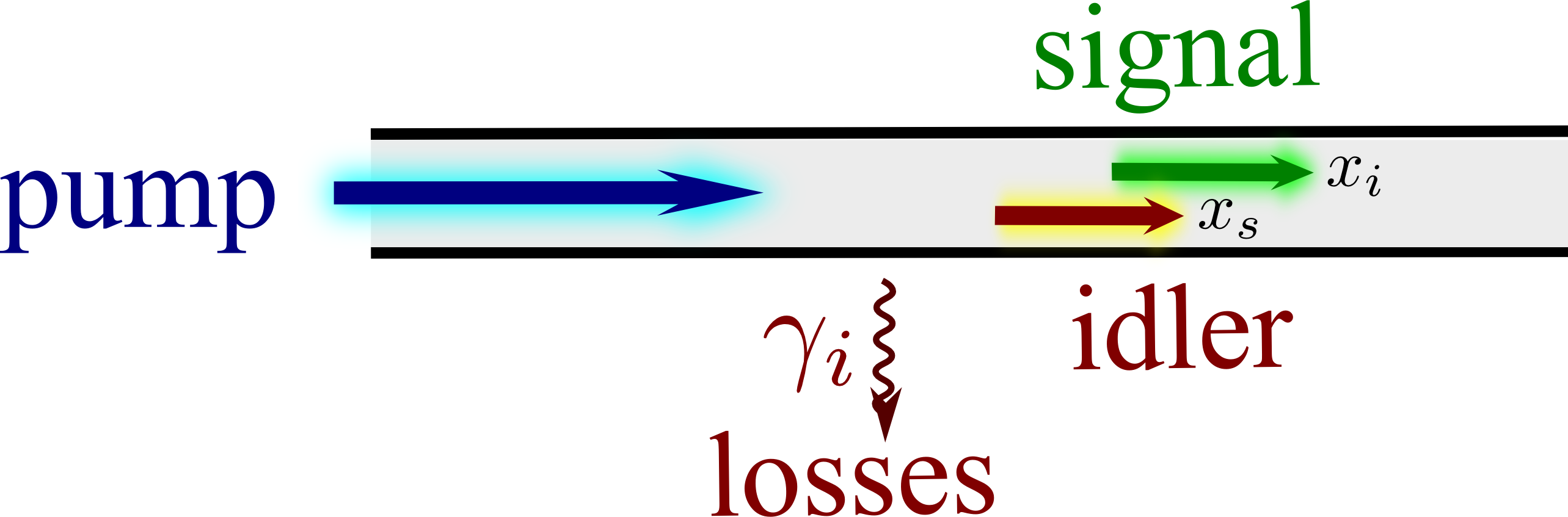}
\end{center}
\caption{Illustration of the two-photon generation in a nonlinear lossy waveguide.}
\label{fig:waveguide}
\end{figure}
\subsection{Correspondence to the  results for waveguides}\label{sec:correspond}
Now we consider an example of one-dimensional dispersionless waveguides with weak losses, see Fig.~\ref{fig:waveguide}. We will
show how  our general Green function theory  allows one to recover  the results known in literature~\cite{Antonosyan:2014-43845:PRA, Helt:2015-13055:NJP}.

The Green function for the one-dimensional scalar problem reads
\begin{equation}
G(x,x')=\frac{\rmi (\omega/c)^2}{2q}\e^{\rmi q|x-x'|}
\label{seq:G1D}
\end{equation}
and satisfies the inhomogeneous Helmholtz equation
\begin{equation}
\left(\frac{\rmd^2}{\rmd x^2}+q^2\right)G(x,x')=-4\pi\left(\frac{\omega}{c}\right)^2\delta(x-x')\:.
\end{equation}
Here, $q$ is the wave propagation constant (photon wave vector), that can be complex for lossy medium. 
The results in Ref.~\cite{Antonosyan:2014-43845:PRA} implied (i) weak losses and (ii) one-way propagation. Using these additional assumptions for the Green function in Eq.~\eqref{seq:G1D} we obtain
\begin{equation}
G(x,x')=\rmi g\e^{\rmi \beta(x-x')-\gamma(x-x')}\theta(x-x'),\quad g=\frac{(\omega/c)^2}{2\beta}\label{seq:G1D2}\:.
\end{equation}
Here, we explicitly introduced the real and imaginary parts of the propagation constant $q=\beta+\rmi \gamma$, neglected the imaginary part of $q$ in the factor $g$ and took into account only the term in the factor $\e^{\rmi q|x-x'|}=\theta(x-x')\e^{\rmi q(x-x')}+\theta(x'-x)\e^{\rmi q(x'-x)}$, corresponding to the waves propagating along the positive direction of the real axis.
We present the generation matrix for the SPDC process as
\begin{equation}
\Gamma(x_0)=\frac{\chi}{2}\mathcal E_p\e^{-\gamma_p x_0}\theta(x-x_0)\:,
\end{equation}
where without the loss of generality we assumed that the real part of the propagation constant for the pump wave is set to zero. 
The two-photon wave function determined by  Eq.~\eqref{seq:Tis113} then reads
\begin{multline}\label{seq:Tis2}
T_{is}(x_{i},\omega_{i};x_{s},\omega_{s})
=-d_i^*d_s^*g_i^* g_s^*\chi\mathcal E_p\times\\\int\limits_{0}^{\min(x_{i},x_{s})} \rmd x_{0}
\e^{(\rmi \beta_{s}-\gamma_{s})(x_{s}-x_{0})}\e^{(\rmi \beta_{i}-\gamma_{i})(x_{i}-x_{0})}\e^{-\gamma_{p}x_{0}}\:.
\end{multline}
For $x_{i}=x_{s}$ this answer for $T(x_{s},x_{s})$ exactly matches the solution in Eq.~(9)
of the differential Eq.~(5)  in Ref.~\cite{Antonosyan:2014-43845:PRA}. 
In order to determine the single-photon count rate we  use Eq.~\eqref{seq:WsAll2} and take into account that for small losses $\Im \eps(\omega_i)$  is proportional to $ \gamma_i$.
Next, we present the integral in Eq.~\eqref{seq:WsAll2} as
\begin{equation}
\gamma_{i}\int \rmd\omega_{i}\int\limits_{0}^{\infty} \rmd x_{i}W_{is}(x_{i},\omega_{i};x_{s},\omega_{s})\:=
\gamma_{i}\int\limits_{0}^{x_{s}} \rmd x_{i}|T_{is}(x_{i},x_{s})|^{2}+
\gamma_{i}\int\limits_{x_{s}}^\infty \rmd x_{i}|T_{is}(x_{i},x_{s})|^{2}\:,\label{seq:two_terms}
\end{equation}
where the two-photon wave function in Eq.~\eqref{seq:Tis2} has the form
\begin{equation}
T_{is}(x_{i},x_{s})=\begin{cases}
\e^{(\rmi \beta_{s}-\gamma_{s})(x_{s}-x_{i})}T_{is}(x_{i},x_{i})
\:,&(x_{i}<x_{s})\:,\\
\e^{(\rmi \beta_{i}-\gamma_{i})(x_{i}-x_{s})}T_{is}(x_{s},x_{s})\:,&(x_{i}>x_{s})\:.\label{seq:two_lines}
\end{cases}
\end{equation}
Substituting Eq.~\eqref{seq:two_lines} into Eq.~\eqref{seq:two_terms} we find 
\begin{equation}
W_s(x_s)\propto 
2\gamma_{i}\int\limits_{0}^{x_{s}} \rmd x_{i}|T_{is}(x_{i},x_{s})|^{2}+
|T_{is}(x_{s},x_{s})|^{2}\label{seq:two_terms2}\:.
\end{equation}
The first term in Eq.~\eqref{seq:two_terms2} exactly corresponds to the term
$\tilde I_{s}(z)$ in Eq.~(10) of Ref.~\cite{Antonosyan:2014-43845:PRA}, i.e. the contribution from the states, where the idler photon has been absorbed. The second term in Eq.~\eqref{seq:two_terms2} matches the term $I_{s}^{(0)}$ in  Eq.~(10) of Ref.~\cite{Antonosyan:2014-43845:PRA}. It is the contribution of  the states where the idler photon 
has still not been absorbed up to the point $x_{s}$. 
\subsection{Application for surface plasmons}\label{sec:plasmons}
\begin{figure}[t]
\begin{center}
\includegraphics[width=0.99\columnwidth]{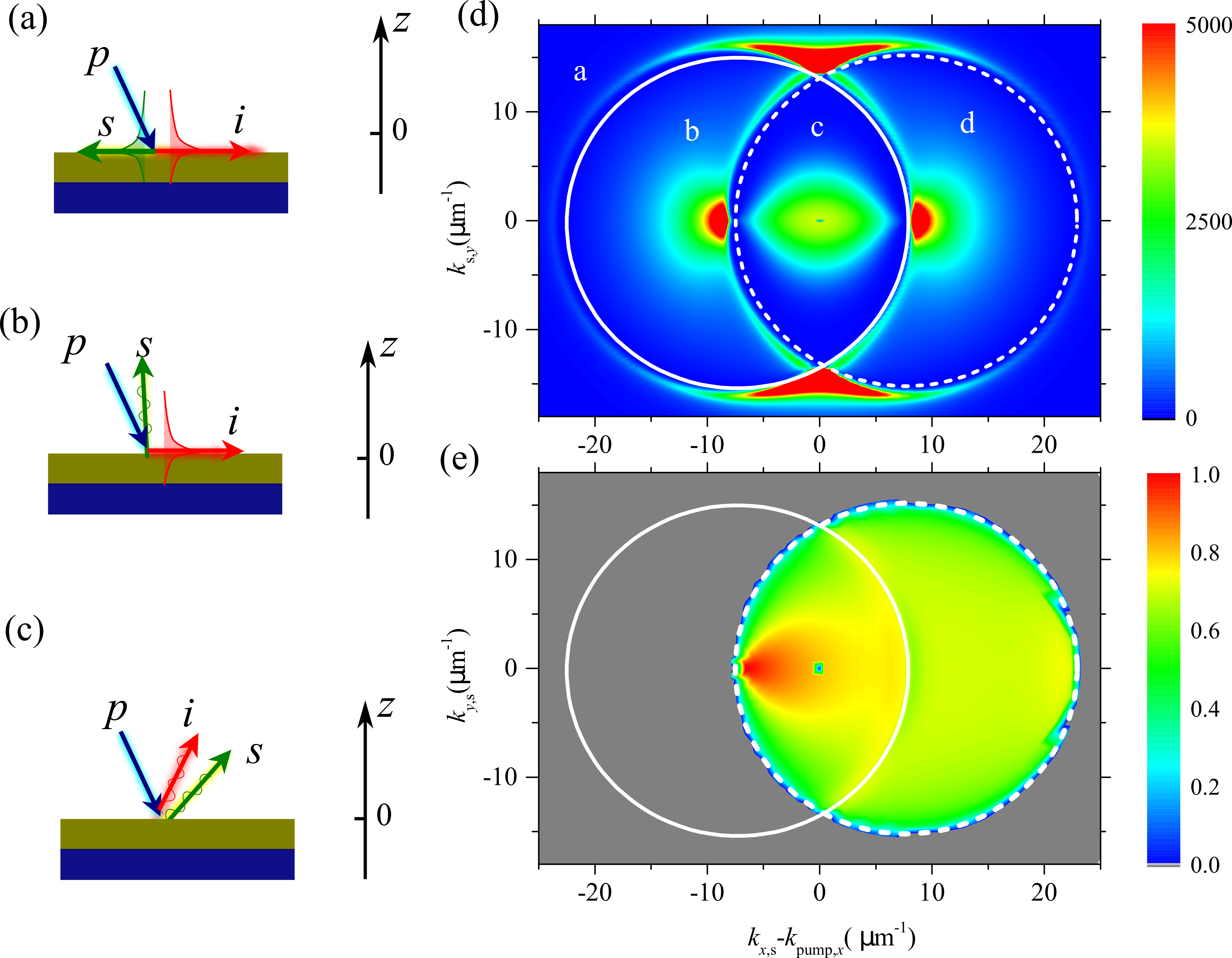}
\end{center}
\vspace*{-5mm}
\caption{\label{fig:2}
(a)-(c) Scheme of SFWM generation of a pair of (a) entangled plasmons, (b) photon entangled with plasmon and (c) entangled photons in the gold/nonlinear dielectric structure.
(d) Color map of the two-photon detection probability $|T(\bm k_{i},\bm k_{s})|^{2}$ in the reciprocal space vs. the in-plane wave vector components (arb.un.) in TM polarization ($\bm d_{i,s}\propto  \bm k\times\bm k\times\hat{\bm z}$) at $z_{i}=z_{s}=100$~nm. The signal (solid) and idler (dashed) light lines are plotted in white color. The letters a--d mark the near- and far-field signal and idler generation regimes.
(e) Efficiency of signal heralding by far field idler photons, Eq.~\eqref{eq:QE}.
For all plots $\hbar\omega_{i}\approx \hbar\omega_{s}\approx \hbar\omega_{p}\approx 3~$eV, $\eps_{\rm diel}=2$, $d_{\rm silver}=20$~nm, pump is TM polarized, $k_{p,x}=0.5\omega_{p}/c$. After Ref.~\cite{Poddubny:2016-123901:PRL}.
}
\end{figure}

We now apply our general theory to
layered metal-dielectric plasmonic structures. First, we analyze the degenerate spontaneous four-wave mixing for the  metallic layer {of the thickness $d_{\rm silver}=20$~nm} on top of the nonlinear dielectric, see Figs.~\ref{fig:2}(a)--(c).
Due to the translational symmetry, the total in-plane momentum $\bm k$ of the photons and plasmons is conserved, i.e.
$k_{i,\alpha}+k_{s,\alpha}=2k_{p,\alpha}$ for $\alpha=x,y$.
The most interesting situation is realized for oblique pump incidence, giving rise to four different regimes when (a) both signal and idler, (b) only idler, (c) neither signal nor idler and (d) only signal in-plane wave vectors lie outside the corresponding light cone boundaries $\omega_{i,s}/c$.
The first three situations are schematically shown in Figs.~\ref{fig:2}(a)--(c). Two-photon generation occurs in case (c), while (b) and (d) correspond to plasmon generation heralded by the far field photon.

We perform numerical simulations considering isotropic dielectric with {electronic} $\chi^{(3)}$ nonlinearity tensor as~\cite{Boyd:2008:NonlinearOptics}: $\chi_{\alpha\beta\gamma\delta}=\chi_{0}(\delta_{\alpha\beta}\delta_{\gamma\delta}
+\delta_{\alpha\delta}\delta_{\beta\delta}+\delta_{\alpha\gamma}\delta_{\beta\delta})$.
We plot the Fourier transform of the two-photon detection amplitude
$|T_{is}(\bm k_{s},z_{i},z_{s})|^{2}$ for $z_{i}=z_{s}=100~$nm above the structure, defined as
$T(\bm k_{i})=\int \rmd x\rmd y \exp(-\rmi k_{x}x-\rmi k_{y}y) T(x,y)$, which characterizes the signal-idler generation efficiency in all different regimes.
The relevant Fourier transforms of the Green functions were evaluated analytically following Ref.~\cite{Tomas:1995-2545:PRA}. {Silver permittivity has been taken from Ref.~\cite{Johnson:1972-4370:PRB} and includes the losses and dispersion.}
The overall map of the correlations resembles that for the generation of the polarization-entangled photons from a bulk nonlinear uniaxial crystal~\cite{Kwiat:1995-4337:PRL}: it shows strong maxima at the intersections of the signal and idler light cone boundaries.
However, contrary to the bulk, the calculated map reflects the two-quantum correlations of both photons and plasmons.  In the region (c) the shown signal can be directly measured from the far field photon-photon correlations.
{For the chosen $30^\circ$ pump incidence  angle the bright spot in the region (b) of  Fig.~\ref{fig:2}(d) corresponds to the signal photons emittered in the normal direction.}
The near-field signal in the regions (a),(b),(d) can be recovered by using the grating to outcouple the plasmons to the far field~\cite{DiMartino:2012-2504:NANL} or with the near field scanning optical microscopy setup~\cite{leFeber:2014-43:NPHOT}. 

The bright spot in the map Fig.~\ref{fig:2}(d) for $k_{s,x}-k_{p,x}\approx 10~\mu$m$^{-1}$ reveals the resonantly enhanced plasmonic emission heralded by the normally propagating idler photons. The heralding efficiency in Eq.~\eqref{seq:E}, adopted for the planar geometry, reads
\begin{align}
QE&=\sum\limits_{z_{i}=-L,L}\sum\limits_{\bm d_{i}=\hat{\bm x},\hat{\bm y},\hat{\bm z}}\frac{c \cos\theta_{i}}{2\pi\hbar\omega_{i}}\frac{|T_{is}(\bm k_{s},z_{s},z_{i},\bm d_{i})|^{2}}{W_{s}(\bm k_{s})}\:,\label{eq:QE}
\end{align}
where
$\cos\theta_{i}=\sqrt{1-(ck_{i}/\omega_{i})^{2}}$.
The summation over $z_{i}$ in Eq.~\eqref{eq:QE} accounts for the total idler photon flux through the surfaces $z_{i}=\pm L$ above and below the nonlinear structure. The calculated values of the signal heralding, shown in Fig.~\ref{fig:2}(e), are remarkably high. They reach almost 100\% in the case when both signal and idler photon are in the far field, see the bright spot at $k_{s,x}-k_{p,x}\approx-5~\mu$m$^{-1}$. In the case of signal plasmons the heralding efficiency is uniform and about $70\%$.
We note that the results in Fig.~\ref{fig:2}(e) correspond to the internal heralding efficiency, calculated for the plane pump wave.
The external quantum heralding efficiency has to account also for the plasmonic losses due to the propagation from the pump spot to the near field detector, which can be optimized in the actual experimental setup.
\section{Quantum-classical correspondence}\label{sec:quantum-classical}
\begin{figure}[t]
\begin{center}
\includegraphics[width=0.75\textwidth]{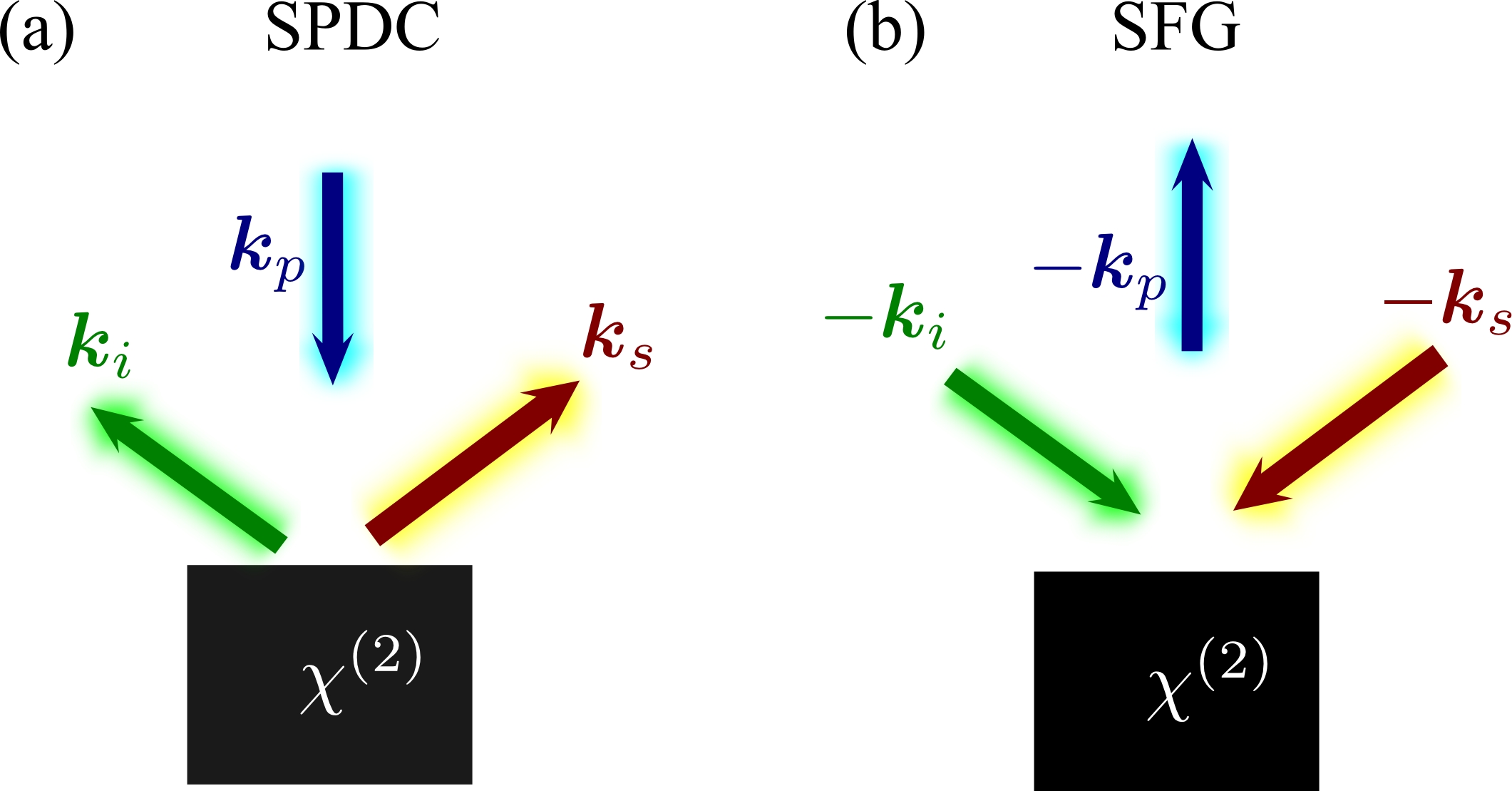}
\end{center}
\caption{Schematic illustration of the reciprocal (a)~spontaneous parametric down-conversion and (b)~sum-frequency generation processes. 
}\label{fig:SuppGeom}
\end{figure}

The characterization of the two-photon state generated by a structured nonlinear system is a hard experimental task~\cite{James:2001-52312:PRA, Altepeter:2005-105:AAMOP, Lvovsky:2009-299:RMP, Titchener:2018-19:NPJQI}. It requires a number of measurements and resources that increase quadratically with system size, making the characterization of states generated by large waveguide circuits impractical or very time consuming.
Additionally, integrated circuits are typically patterned on wafers and produced in large numbers, and efficient techniques for fast quality checking of device performance are urgently needed. Here we present an efficient method for the characterization of two-photon states generated from arbitrary optical structures with quadratic nonlinearity that has both fundamental and practical importance for the development of future integrated quantum photonics technologies.

A practical approach for predicting the biphoton state produced from a nonlinear device using only classical detectors and laser sources was proposed in Ref.~\cite{Liscidini:2013-193602:PRL} based on the concept of Stimulated emission tomography (SET).
This technique takes advantage of the analogy between a spontaneous nonlinear process and its classical stimulated counterpart, i.e. difference-frequency generation or stimulated four-wave mixing. It was applied experimentally for spectral characterization of two-photon states with an accuracy unobtainable with single-photon detection methods~\cite{Eckstein:2014-L76:LPR, Fang:2014-281:OPT,Jizan:2015-12557:SRP, Grassani:2016-23564:SRP}, and fast reconstruction of the density matrix of entangled-photon sources~\cite{Rozema:2015-430:OPT, Fang:2016-10013:OE}. 

However, for large optical networks SET becomes a challenging task as discussed in Ref.~\cite{Titchener:2015-33819:PRA}, which prevented its experimental realizations for multi-mode path entangled states. The reason is that one would need to precisely inject the seed beam in the individual supermodes supported by the structure. A possible workaround is to inject the seed beam in each single channel individually and to perform a transform through supermode decomposition to obtain quantum predictions. In either case, one requires complete knowledge of the linear light dynamics inside the whole structure, making SET a multi-step procedure prone to errors and not applicable to ``black-box'' circuits. Additionally, the analogy between a spontaneous nonlinear process and its stimulated counterpart is strictly valid only in the limit of zero propagation losses~\cite{Helt:2015-1460:OL}. This assumption poses a fundamental limit for the characterization of integrated waveguide circuits. Indeed, the effect of scattering losses, due to impurities or surface and side-wall roughness, becomes prominent with increasing miniaturization of photonic devices.
Sum-frequency generation (SFG), the reverse process of SPDC, was identified in Ref.~\cite{Helt:2015-1460:OL} as the ideal approach for characterizing  second-order nonlinear circuits in presence of losses. Nevertheless, the method was formulated only for a single and homogeneous waveguide, posing a stringent restriction for the characterization of more complex devices. 

Here, we overcome the limitations of the previous proposals by establishing a rigorous equivalence between the biphoton wavefunction in the undepleted pump regime and the sum-frequency field generated by classical wave-mixing in the reverse direction of SPDC. Our theory analysis is generalized by the use of the Green-function method~\cite{Poddubny:2016-123901:PRL,Lenzini:2018-17143:LSA}, and holds for arbitrarily complex second-order nonlinear circuits and in presence of propagation losses. More importantly, the SFG-SPDC analogy can be expressed in any measurement basis, providing a simple experimental tool for the characterization of any ``black-box'' $\chi^{(2)}$-nonlinear process, see Fig.~\ref{fig:SuppGeom}.
\subsection{Quantum SPDC-SFG reciprocity relationship}
We start with proving the general identity for the correspondence between the SPDC process and the sum-frequency generation (SFG) processes in the reversed geometry. This identity will    generalize  the Lorentz reciprocity theorem in linear electromagnetism~\cite{BornWolf:1999} 
\begin{equation}
\int\rmd^{3}r\bm P_{1}(\bm r)\cdot \bm E_{2}(\bm r)=
\int\rmd^{3}r\bm P_{2}(\bm r)\cdot \bm  E_{1}(\bm r)\label{eq:Lorentz}
\end{equation}
that links the electric field distribution  $\bm E_{1}(\bm r)$ and $\bm E_{2}(\bm r)$, induced by the polarization distributions $\bm P_{1}$ and $
\bm P_{2}$, respectively.
The biphoton wavefunction  in the SPDC regime, following from Eq.~\eqref{seq:Tis113}, reads
\begin{equation}\label{eq:Tsi1}
\Psi(\bm r_{s},\bm r_{i},\sigma_{s},\sigma_{i},\omega_{s},\omega_{i})=\int\rmd^{3} r_{0}
G_{\sigma_{s}\alpha}(\bm r_{s},\bm r_{0};\omega_{s})
G_{\sigma_{i}\beta}(\bm r_{i},\bm r_{0};\omega_{i})\chi^{{(2)}}_{\alpha\beta\gamma}E_{p,\gamma}(\bm r_{0})\:.
\end{equation}
On the other hand, the nonlinear wave at the sum frequency $\omega_{p}=\omega_{i}+\omega_{s}$ generated from the waves $\bm E_{s}(\bm r_{i})$, $\bm E_{i}(\bm r_{i})$ in the nonlinear structure can be presented as
\begin{equation}\label{eq:ESFG1}
E_{SFG,\sigma_{p}}(\bm r_{p})=\int \rmd ^{3}r_{0} G^{\vphantom{(2)}}_{\sigma_{p}\gamma}(\bm r_{p},\bm r_{0}) \chi^{(2)}_{\alpha\beta\gamma}(\bm r_{0})E_{s,\alpha}(\bm r_0)E_{i,\beta}(\bm r_0)\:.
\end{equation}
Inspired by linear reciprocity relationship Eq.~\eqref{eq:Lorentz}  we introduce the polarizations $P_{i,s,p}(\bm r)$ inducing the correspondent waves $E_{s,i,p}(\bm r)$,
\begin{equation}
\bm E_{\nu}(\bm r)=\int\rmd^{3}r_{0}\hat G(\bm r,\bm r_{0};\omega_{p})\bm P_{\nu}(\bm r_{0}),\quad \nu=i,s,p\:.
\end{equation}
This allows us to rewrite Eq.~\eqref{eq:Tsi1} and Eq.~\eqref{eq:ESFG1} as
\begin{multline}\label{eq:Tsi2}
\Psi(\bm r_{s},\bm r_{i},\sigma_{s},\sigma_{i},\omega_{s},\omega_{i})=\int\rmd^{3} r_{0}
\int\rmd^{3} r_{p}
G_{\sigma_{s}\alpha}(\bm r_{s},\bm r_{0})
G_{\sigma_{i}\beta}(\bm r_{i},\bm r_{0})\chi^{{(2)}}_{\alpha\beta\gamma}(\bm r_{0})\times\\G_{\gamma \sigma_{p}}(\bm r_{0},\bm r_{p})P_{p,\sigma_{p}}(\bm r_{p})\:,
\end{multline}
and 
\begin{multline}\label{eq:ESFG2}
E_{SFG,\sigma_{p}}(\bm r_{p};\omega_{i}+\omega_{s})=\int \rmd ^{3}r_{0}\int \rmd ^{3}r_{i}\int \rmd ^{3}r_{s} G^{\vphantom{(2)}}_{\sigma_{p}\gamma}(\bm r_{p},\bm r_{0}) \chi^{(2)}_{\alpha\beta\gamma}(\bm r_{0})\times\\G_{\alpha\sigma_{s}}(\bm r_0,\bm r_{s})G_{\alpha\sigma_{i}}(\bm r_0,\bm r_{i})
E_{s,\sigma_{s}}(\bm r_{s})E_{i,\sigma_{i}}(\bm r_{i})
\:.
\end{multline}
We have omitted the frequency arguments in the Green functions for the sake of brevity. In the reciprocal structure the Green functions satisfy the reciprocity property 
\begin{equation}\label{eq:Gab12}
G_{\alpha\beta}(\bm r_{1},\bm r_{2})=G_{\beta\alpha}(\bm r_{2},\bm r_{1})\:,
\end{equation}
that is equivalent to Eq.~\eqref{eq:Lorentz}. 
Comparing Eq.~\eqref{eq:Tsi2} and Eq.~\eqref{eq:ESFG2} with the help of Eq.~\eqref{eq:Gab12} we establish the general reciprocity relationship  between SPDC and SFG processes in the form
\begin{multline}\label{eq:gen-correspondence}
\iint \rmd^{3}r_{i}\rmd^{3}r_{s}
\Psi(\bm r_{s},\bm r_{i},\sigma_{s},\sigma_{i},\omega_{s},\omega_{i})P_{i,\sigma_{i}}(\bm r_{i})P_{s,\sigma_{s}}(\bm r_{i})\\=
\int \rmd^{3}r_{p}E_{SFG,\gamma}(\bm r_{p};\omega_{i}+\omega_{s})P_{p, \gamma}(\bm r_{p})\:.
\end{multline}
\subsection{SPDC-SFG correspondence for a localized nonlinear source}
Now we apply the general SPDC-SFG reciprocity relationship Eq.~\eqref{eq:gen-correspondence} to the case of photon pair generation 
from a localized nonlinear source. 
We start with calculating electric field of the structure illuminated by two plane waves,
the ``idler'' one with the wave vector $-\bm k_{i}$ and the ``signal'' one with the wave vector $-\bm k_{s}$, see Fig.~\ref{fig:SuppGeom}(b).
To this end we assume that both waves are generated by point dipoles located in the far field zone in the points
$\bm r_{i,s}\parallel \bm k_{i,s}$,
and having the unit amplitudes $\bm d_{i}^{\ast} \perp \bm k_{i}$ and $\bm d_{s}^{\ast}\perp \bm k_{s}$, respectively.
In this case the linear fields at idler and signal frequencies can be written as
\begin{equation}
\bm E_{i,s}(\bm r)=G(\bm r,\bm r_{i,s};\omega_{i,s})\bm d_{i,s}^{\ast}\label{eq:Eis}
\end{equation}
where $G$ is the electromagnetic Green function tensor satisfying the equation
\begin{equation}
\rot\rot G(\bm r,\bm r';\omega)=\left(\frac{\omega}{c}\right)^{2}\eps(\bm r)G(\bm r,\bm r';\omega)+4\pi 
\left(\frac{\omega}{c}\right)^{2}\delta(\bm r-\bm r')\:.
\end{equation}
Far away from the nonlinear structure, where $\eps=1$, the Green function reduces to the free Green function
\begin{equation}
G_{\alpha\beta}=\left[\left(\frac{\omega}{c}\right)^{2}+\frac{\partial^{2}}{\partial x_{\alpha}\partial x_{\beta}}\right]
\frac{\e^{\rmi \omega |\bm r-\bm r'|/c}}{|\bm r-\bm r'|}\:.\label{eq:G0}
\end{equation}
The locally-plane idler and signal waves, incident upon the structure, are found by substituting
Eq.~\eqref{eq:G0} into Eq.~\eqref{eq:Eis} and assuming $r_{i,s}\gg r$.
Hence, we find 
\begin{equation}
\bm E_{i,s}^{(0)}=\frac{\e^{\rmi \omega_{i,s}r_{i,s}/c}}{r_{i,s}} q_{i,s}^{2}\bm d_{i,s}^{\ast}\:,\label{eq:E0is}
\end{equation}
where $q_{i,s}=\omega_{i,s}/c$. The corresponding time-averaged fluxes are found as 
\begin{equation}
\Phi_{i,s}=\frac{c}{2\pi}|E_{i,s}^{(0)}|^{2}=\frac{c}{2\pi}\frac{q_{i,s}^{4}}{r_{i,s}^{2}}|d_{i,s}|^{2}\:.\label{eq:Phiis}
\end{equation}
The nonlinear SFG field is found as a convolution of the $\chi^{(2)}$ susceptibility, the incident fields, and the Green function at the sum frequency:
\begin{multline}
E_{\alpha}^{(\rm SFG)}(\bm r_{p}\leftarrow \bm r_{i},\bm d_{i}^{\ast};\bm r_{s},\bm d_{s}^{\ast})\\
= 
\int \rmd^3 r' G_{\alpha\beta }(\bm r_{p},\bm r')
\chi^{(2)}_{\gamma\delta,\beta}G_{\gamma\nu }(\bm r',\bm r_{i}) G_{\delta\mu }(\bm r',\bm r_{s})d_{i,\nu}^{\ast}d_{s,\mu}^{\ast}\:.
\label{eq:ESFG}
\end{multline}
Here $\alpha,\beta,\gamma,\delta,\mu,\nu$ are the Cartesian indices and the frequency arguments of the Green functions are omitted for the sake of brevity.
It is instructive to present the Green functions in the following way
\begin{equation}
G_{\alpha\beta }(\bm r,\bm r')=q^{2}\frac{\e^{\rmi qr}}{r} g_{\alpha\beta }\left(\tfrac{\bm r}{r},\bm r'\right) \text{ for } r\gg c/\omega, r\gg r'\:,
\label{eq:gsmall}
\end{equation}
where the dimensionless scattering amplitudes $g_{\alpha\beta }(\frac{\bm r}{r},\bm r')\equiv 
g_{\alpha\beta }(\bm k,\bm r')$ describe the conversion between the near field at the point $\bm r'$ and the plane wave propagating in the direction $\bm r/r\parallel \bm k$. We also use the Lorentz reciprocity property
\begin{equation}G_{\alpha\beta }(\bm r,\bm r')=G_{\beta\alpha }(\bm r',\bm r)
\label{eq:Lorentz2}
\end{equation}
Applying Eq.~\eqref{eq:gsmall} and  Eq.~\eqref{eq:Lorentz2} to Eq.~\eqref{eq:ESFG} we rewrite the sum-frequency wave as
\begin{multline}
E_{\alpha}^{(\rm SFG)}(\bm r_{p}\leftarrow \bm r_{i},\bm d_{i}^{\ast};\bm r_{s},\bm d_{s}^{\ast})=\\\frac{q_{i}^{2}q_{s}^{2}q_{p}^{2}}{r_{i}r_{s}r_{p}}
\e^{\rmi (q_{i}r_{i}+q_{s}r_{s}+q_{p}r_{p})}\int  \rmd^3 r' g_{\alpha\beta }(\bm k_{p},\bm r')
\chi^{(2)}_{\gamma\delta,\beta}g_{\nu\gamma }(\bm k_{i},\bm r')
g_{\mu\delta }(\bm k_{s},\bm r')
d_{i,\nu}^{\ast}d_{s,\mu}^{\ast}\:.\label{eq:ESFG3}
\end{multline} 
Now we introduce the  differential SFG efficiency   as
\begin{equation}
\rmd \Xi^{\rm SFG}(-\bm k_{i},\bm e^{\ast}_{i};-\bm k_{s}\bm e_{s}^{\ast}\to -\bm k_{p},\bm e^{\ast}_{p})=
r_{p}^{2}\rmd \Omega_{p}\frac{\Phi_{p}( -\bm k_{p},\bm e^{\ast}_{p})}{
\Phi_{i}( -\bm k_{i},\bm e^{\ast}_{i})\Phi_{s}( -\bm k_{s},\bm e^{\ast}_{s})}\:,\label{eq:sigma}
\end{equation}
where $\bm e_{s,i} = \bm d_{s,i} / |\bm d_{s,i}|$. The quantity in Eq.~\eqref{eq:sigma} represents the ratio of the  power of SFG photons propagating inside the solid angle $\rmd \Omega_{p}$ in the direction 
$-\bm k_{i}$ to the energy fluxes of incoming signal and idler plane waves $\Phi_{i}$ and $\Phi_{s}$.  
Eq.~\eqref{eq:sigma} bears analogies with the  scattering cross section in the linear problem.
The value of SFG efficiency is found from Eq.~\eqref{eq:ESFG} and Eq.~\eqref{eq:Phiis} 
as 
\begin{multline}
\frac{\rmd \Xi^{\rm SFG}(-\bm k_{i},\bm e^{\ast}_{i};-\bm k_{s}\bm e_{s}^{\ast}\to -\bm k_{p},\bm e^{\ast}_{p})}{\rmd \Omega_{p}}=\frac{2\pi q_{p}^{4}}{c}\times\\\left|\int  \rmd^3 r' e_{p,\alpha}g_{\alpha\beta }(\bm k_{p},\bm r')
\chi^{(2)}_{\gamma\delta,\beta}g_{\nu\gamma }(\bm k_{i},\bm r')
g_{\mu\delta }(\bm k_{s},\bm r')
e_{i,\nu}^{\ast}e_{s,\mu}^{\ast}\right|^{2}\:.\label{eq:sigma2}
\end{multline} 

Now we proceed to the SPDC process. The complex  wavefunction of a photon pair, generated in a $\chi^{(2)}$-nonlinear structure, has the amplitude~\cite{Poddubny:2016-123901:PRL}
\begin{equation}
T(\bm r_{s}\mu,\bm r_{i}\nu\leftarrow \bm r_{p}\bm e_{p})=\int \rmd^{3} r'
G_{\sigma_{s}\nu}(\bm r_{s},\bm r')G_{\sigma_{i}\nu}(\bm r_{i},\bm r')
\chi^{(2)}_{\gamma\delta,\beta}(\bm r')E_{p,\beta}(\bm r',\omega_{p})\:,\label{eq:T-SPDC}
\end{equation}
where $\bm r_{s} (\bm r_{i})$ and $\mu (\nu)$ are signal (idler) photon coordinates and polarizations, respectively, and $E_{p}$ is the electric field of the pump with the frequency $\omega_{p}$.  Similar to Eqs.~\eqref{eq:Eis} in the SFG case we now write that the pump wave is generated by the point far-field source,
\begin{equation}
\bm E_{p}(\bm r)=G(\bm r,\bm r_{p})\bm d_{p}\label{eq:Ep}
\end{equation}
where $\bm d_{p}\parallel \bm e_{p}\perp \bm r_{p}$ is the point dipole amplitude.
Plugging in the Green function asymptotic expressions from Eq.~\eqref{eq:gsmall}, we rewrite the biphoton wavefunction in the form
\begin{multline}
T(\bm r_{s}\mu,\bm r_{i}\nu\leftarrow \bm r_{p}\bm e_{p})=
\frac{q_{i}^{2}q_{s}^{2}q_{p}^{2}}{r_{i}r_{s}r_{p}}
\e^{\rmi (q_{i}r_{i}+q_{s}r_{s}+q_{p}r_{p})}\\\times\int  \rmd^3 r' g_{\alpha\beta }(\bm k_{p},\bm r')
\chi^{(2)}_{\gamma\delta,\beta}g_{\nu\gamma }(\bm k_{i},\bm r')
g_{\mu\delta }(\bm k_{s},\bm r')
d_{p,\alpha}\:.\label{eq:T-SPDC2}
\end{multline}
Comparing Eq.~\eqref{eq:T-SPDC2} with Eq.~\eqref{eq:ESFG3} we find the identity
\begin{equation}
T(\bm r_{s}\mu,\bm r_{i}\nu\leftarrow \bm r_{p}\bm e_{p})d^{\ast}_{s,\mu}d^{\ast}_{i,\nu}=
E_{\alpha}^{(\rm SFG)}(\bm r_{p}\leftarrow \bm r_{i},\bm d_{i}^{\ast};\bm r_{s},\bm d_{s}^{\ast})d_{p,\alpha}\:,
\end{equation}
which proves the  SPDC-SFG correspondence for the localized $\chi^{(2)}$-nonlinear source.

We are also interested in comparing the experimentally accessible quantities, namely, the two-photon pair generation rate and the SFG generation efficiency in Eq.~\eqref{eq:sigma2}. 
In order to determine the photon  pair generation rate, we need to calibrate the photon detection process~\cite{Poddubny:2016-123901:PRL, Lenzini:2018-17143:LSA}. To this end we explicitly introduce the signal and idler detectors modelled as the two-level systems with the dipole momenta matrix elements $\bm d_{i}$, $\bm d_{s}$ and the energies $\hbar \omega_{i}$, $\hbar \omega_{s}$. The number of photons absorbed by the detector per unit time is given by
\begin{equation}
\frac{\rmd N_{\rm abs}}{\rmd t}=\frac{2\pi}{\hbar}\delta(\hbar\omega-\hbar\omega_{i,s})|\bm d\cdot \bm E_{i,s}|^2,
\end{equation}
where $\bm E_{i,s}\propto 1/r_{i,s}$ is the local electric field of emitted signal or idler photon at the corresponding detector. 
The detector quantum efficiency $\rmd QE_{i,s}/\rmd \Omega_{i,s}$ is  the ratio between the number of photons absorbed by the detector and the number of photons $\rmd N_{i,s}/\rmd t$ propagating inside the solid angle $\rmd \Omega_{i,s}$ per unit time, 
\begin{equation}
\frac{\rmd N_{i,s}}{\rmd t}=r^{2}\rmd \Omega_{i,s}\frac{c}{2\pi \hbar\omega_{i,s}}|E|^{2}\:,
\end{equation}
\begin{equation}
\frac{\rmd QE_{i,s}}{\rmd \Omega_{i,s}}=\frac{\rmd N_{\rm abs}}{\rmd N_{i,s}}=\frac{4\pi \omega |d_{i,s}|^{2}}{\hbar c}\frac1{r_{i,s}^{2}}\label{eq:Qis}\:.
\end{equation}
The two-photon generation rate per unit of the signal and idler spectra is formally defined as
\begin{equation}
\frac{\rmd N_{\rm pair}}{\rmd t \rmd \omega_i\rmd \omega_s \rmd \Omega_{i}\rmd \Omega_{s}}=\frac{W_{is}}{\rmd QE_i\rmd QE_s}
\label{eq:Npair}
\end{equation}
where
\begin{equation}
W_{is}=\frac{2\pi}{\hbar}\delta(\hbar\omega_p-\hbar\omega_i-\hbar\omega_s)|\sum \limits_{\nu \mu}d_{i,\nu}^{\ast}d_{s,\mu}^{\ast}T(\bm r_{s}\mu,\bm r_{i}\nu\leftarrow \bm r_{p}\bm e_{p})|^2,\label{eq:Wis}
\end{equation}
is the uncalibrated rate of two photon counts calculated from the bi-photon amplitude Eq.~\eqref{eq:T-SPDC2}.
Substituting Eqs.~\eqref{eq:T-SPDC2},\eqref{eq:Qis},\eqref{eq:Npair} into Eq.~\eqref{eq:Wis} and comparing with Eq.~\eqref{eq:sigma2} we find  a general {\it absolute} correspondence between the sum frequency rate  and the photon pair generation rate in the form 
\begin{multline}
\frac{1}{\Phi_{p}}\frac{\rmd N_{\rm pair}(\bm k_{i},\bm e_{i};\bm k_{s}\bm e_{s}\leftarrow \bm k_{p},\bm e_{p})}{\rmd t\rmd \Omega_{i}\rmd \Omega_{s}\rmd \omega_{i}\rmd\omega_{s}}=\\\frac{\delta(\omega_{i}+\omega_{s}-\omega_{p})}{2\pi}
\frac{\lambda_{p}^{4}}{\lambda_{i}^{3}\lambda_{s}^{3}}\frac{\rmd \Xi^{\rm SFG}(-\bm k_{i},\bm e^{\ast}_{i};-\bm k_{s}\bm e_{s}^{\ast}\to -\bm k_{p},\bm e^{\ast}_{p})}{\rmd \Omega_{p}}\:.\label{eq:corr}
\end{multline}
Equation~\eqref{eq:corr} is the main result for the SPDC-SFG correspondence, valid for an arbitrary localized $\chi^{(2)}$-nonlinear system. In order to facilitate comparison with actual experimental setup we first integrate it over the signal and idler frequencies and obtain 
\begin{equation}
\frac{1}{\Phi_{p}}\frac{\rmd N_{\rm pair}(\bm k_{i},\bm e_{i};\bm k_{s}\bm e_{s}\leftarrow \bm k_{p},\bm e_{p})}{\rmd t\rmd \Omega_{i}\rmd \Omega_{s}}=\frac{\Delta \omega_{s}}{2\pi}
\frac{\lambda_{p}^{4}}{\lambda_{i}^{3}\lambda_{s}^{3}}\frac{ \rmd \Xi^{\rm SFG}(-\bm k_{i},\bm e^{\ast}_{i};-\bm k_{s}\bm e_{s}^{\ast}\to -\bm k_{p},\bm e^{\ast}_{p})}{\rmd \Omega_{p}}\:,\label{eq:corr2}
\end{equation}
where $\Delta\omega_{s}\equiv 2\pi c\Delta\lambda_{s}/\lambda_{s}^{2}$ is the signal spectral width. 
\begin{figure}[t!]
\begin{center}
\includegraphics[width=\textwidth]{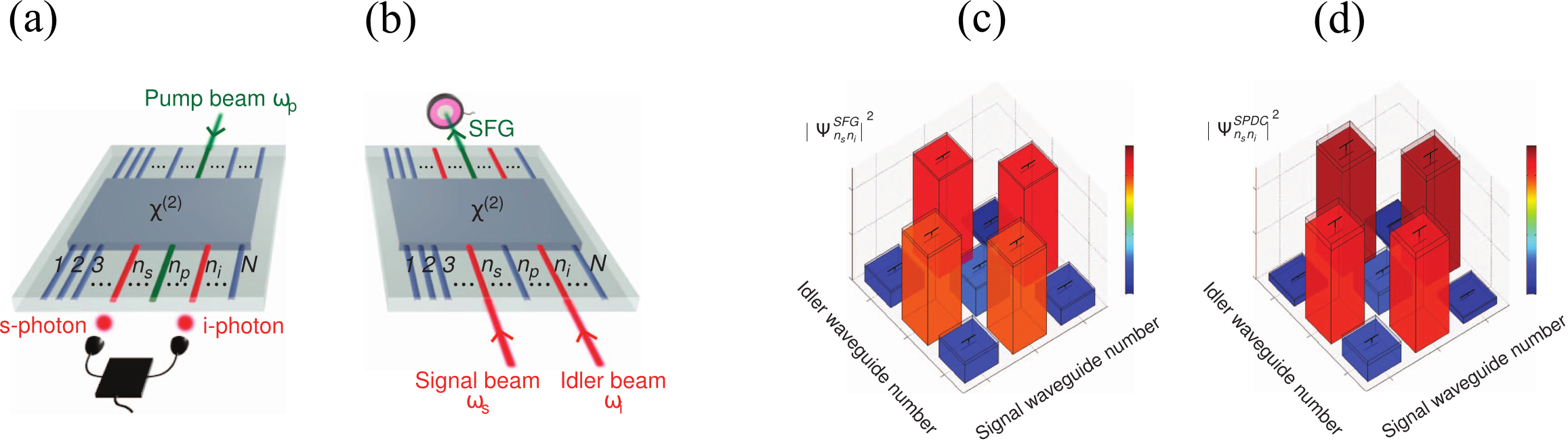}
\end{center}
\caption{Scheme for the characterization of the biphoton state produced by an array of $N$ waveguides with an arbitrary $\chi^{(2)}$-nonlinear process. (a) SPDC: a pump beam is injected into waveguide at the input of the device. Photon-coincidence counting measurements between each pair of waveguides at the output are used to measure photon-pair generation rates and relative absolute squared values of the wavefunction. (b) SFG: Laser light at signal and idler frequencies is injected into waveguides and  in the reverse direction of SPDC. Absolute photon-pair generation rates and relative absolute squared values of the wavefunction can be predicted by direct optical power detection of the sum-frequency field emitted from waveguide $n_{p}$.
(c,d) Normalized biphoton wavefunctions predicted by SFG (c) and measured by SPDC (d). After Ref.~\cite{Lenzini:2018-17143:LSA}.
}\label{fig:Lenzini}
\end{figure}

\subsection{Experimental demonstration for a coupled waveguide system}\label{sec:SPDCexp}
Here we overview the experiment on observation of SPDC-SFG correspondence made in Ref.~\cite{Lenzini:2018-17143:LSA} for the system with coupled nonlinear waveguides.

The SFG-SPDC characterization protocol was realized  for an array of coupled nonlinear waveguides, representing a practical example of complex multi-dimensional system. The measurement schemes for  SPDC  and SFG are shown  in Fig.~\ref{fig:Lenzini}(a) and  Fig.~\ref{fig:Lenzini}(b), respectively. The device is made of three evanescently coupled  waveguides, fabricated on a $z$-cut lithium niobate substrate by the use of the reverse proton exchange technique~\cite{Lenzini:2015-1748:OE, Korkishko:1998-1838:JOSA}. The three waveguides have an inhomogeneous and asymmetric poling pattern along the propagation direction with five defects at different locations of the array introduced by translating the poled domains by half a poling period $\Lambda$. This design is based on the recently developed concept of quantum state engineering with specialized poling patterns~\cite{Titchener:2015-33819:PRA}.

The squared amplitudes of the wavefunction elements predicted by SFG measurements are shown in Fig.~\ref{fig:Lenzini}(c) and those directly measured through SPDC coincidences are presented in Fig.~\ref{fig:Lenzini}(d). The SFG predictions are obtained by integrating the measured conversion efficiencies over a bandwidth of 6~nm. The two correlation matrices have a fidelity of $F=\sum_{n_s n_i} \sqrt{|\Psi_{n_s n_i}^{SFG}|^2 |\Psi_{n_s n_i}^{SPDC}|^2}= 99.28 \pm 0.3\:\%$.

In case of the waveguide geometry, Eq.~\eqref{eq:corr2} for the correspondence between the absolute photon-pair generation rates for SPDC can be written as 
\begin{equation}
\frac{1}{P_p} \frac{{\rm d}N_{\rm pair}}{{\rm d} \omega_s {\rm d}t} =  \frac{\omega_i \omega_s}{2\pi\omega_p^2} \eta^{\rm SFG}_{n_s n_i} (\omega_s,\omega_i)\:.
\label{eq:correspond-wave}
\end{equation}
Here, $P_p$ is the power of the pump beam during SPDC, ${\rm d}N_{\rm pair}/{\rm d} \omega_s {\rm d}t$ is the rate of photon-pair coincidence counts per unit signal frequency, and $\eta^{\rm SFG}_{n_s n_i}\equiv P_{\rm SFG}/(P_{s}P_{i})$ is the sum-frequency power conversion efficiency. Using the SFG measurements and Eq.~\eqref{eq:correspond-wave} a photon pair generation rate has been found as $N_{SFG} = 2.36 \pm 0.14$~MHz, which is the sum of the rates from all 6 output combinations. Direct measurement of this rate from the SPDC data gives $N_{SPDC} = 1.67 \pm 0.15~\rm MHz$, showing a good qualitative agreement between the two values. 
\section{Experimental demonstration of photon-pair generation in dielectric nanoantennas}
\label{sec:experiment}

\begin{figure}[tb!]
    \centering
    \includegraphics[width=0.9\columnwidth]{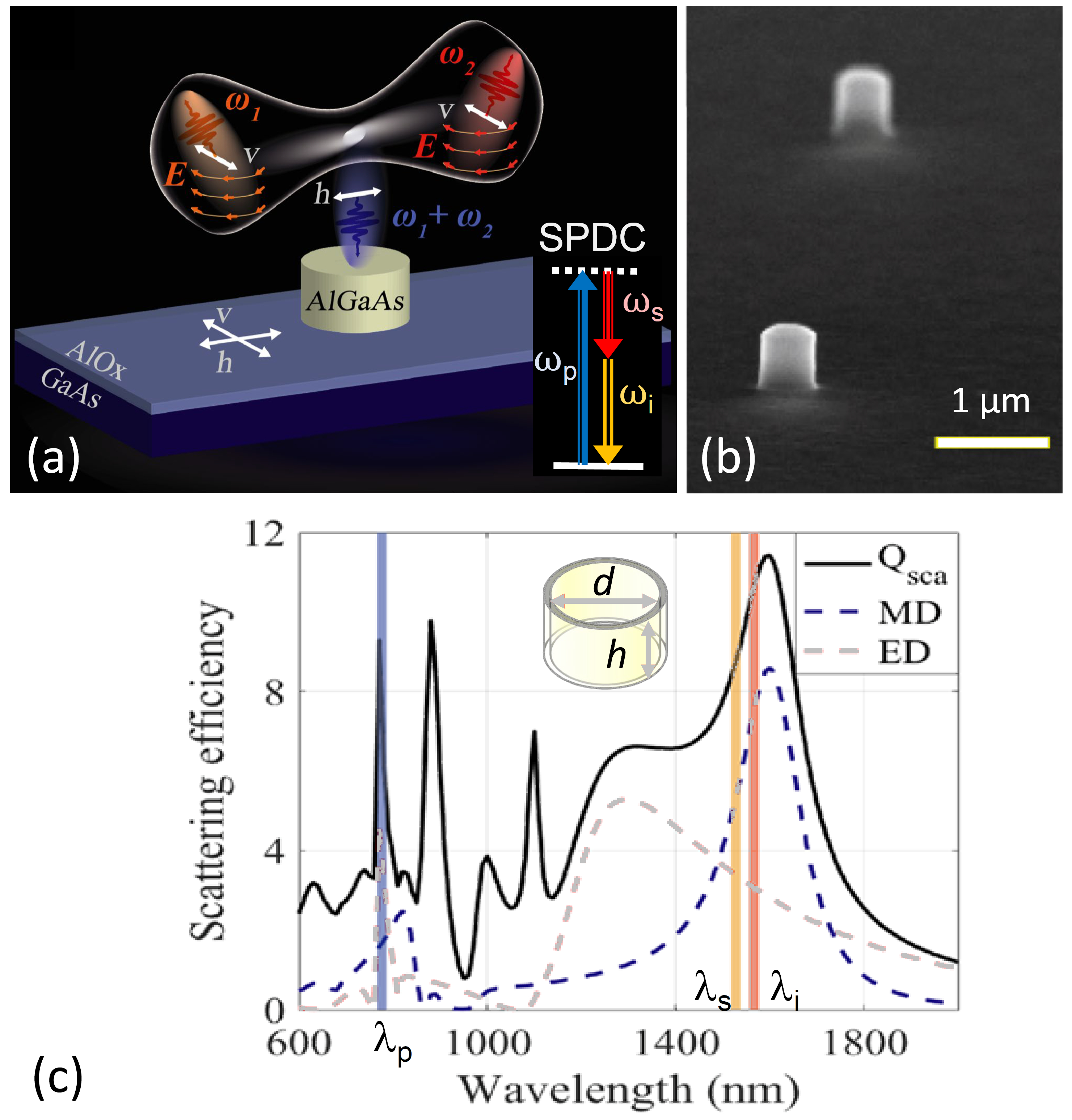}
    \caption{Nonlinear nanoantenna for generation of spontaneous photon pairs.
    (a) Schematic representation of the nanoantenna-based source of photon-pairs through the SPDC process. The inset depicts the energy diagram of the SPDC process. The SPDC pump is horizontally polarized along the $(100)$ AlGaAs crystallographic axis. 
    (b)~A typical scanning electron microscope (SEM) image of two $(100)$ AlGaAs monolithic nanocylinders, $10\,\mu$m apart, such that each cylinder can be excited individually. 
    (c)~Simulated scattering efficiency, $Q_{sca}$, and multipolar decomposition in terms of the two leading electric (ED) and magnetic dipoles (MD) for a nanocylinder with a diameter $d=430$~nm and height $h=400$~nm. The vertical blue and orange lines show the spectral ranges of the pump light and the generated SPDC light (signal and idler), as indicated by labels. The inset shows the geometry of the nanoantenna. After Ref.~\cite{Marino:2019-1416:OPT}.}
\label{fig_concept}
\end{figure}

The experimental demonstration of generation of spontaneous photon pairs in nanoscale photonic structures has been recently demonstrated using an AlGaAs disk nanoantenna exhibiting Mie-type resonances at both pump and bi-photon wavelengths~\cite{Marino:2019-1416:OPT}. 
A schematic of the nanoantenna is shown in Fig.~\ref{fig_concept}(a). It is a crystalline AlGaAs cylinder with a diameter $d=430$~nm and height $h=400$~nm. A scanning electron microscope (SEM) image of the fabricated structure is shown in Fig.~\ref{fig_concept}(b). The non-centrosymmetric crystalline structure of the $(100)$-grown AlGaAs offers strong bulk quadratic susceptibility of $d_{14}=100$~pm/V~\cite{Shoji:1997-2268:JOSB,Ohashi:1993-596:JAP}. The AlGaAs also exhibits high transparency in a broad spectral window from 730~nm up to the far infra-red, due to its direct electronic bandgap. As such, the one- and two-photon absorption at telecommunication wavelengths is negligible. The distance between neighbouring nanocylindors is $10~\mu$m, thereby the response is dominated by the local optical properties of the single antenna~\cite{Gili:2016-15965:OE}. 

The antenna was excited by a linearly polarized pump beam in the near-infrared spectral range and through the process of SPDC generates signal and idler photons in the telecommunication wavelength range. The dimensions of the nanocylinder are chosen such that it exhibits Mie-type resonances at the pump and signal/idler wavelengths. The simulated linear scattering efficiency is defined as the scattering cross section $C_{sca}$ normalized by the cross area of the nanocylinder $\pi r^2$: $Q_{sca} = C_{sca}/\pi r^2$. It is shown in Fig.~\ref{fig_concept}(c) along with the two leading multipolar contributions of the scattering. In the infrared region of the spectrum, where the signal and idler photons are generated, the nanocylinder exhibits a magnetic dipolar resonance, which is the lowest order Mie-mode, featuring a Q-factor of nine [Fig.~\ref{fig_concept}(c)]. For the spectral region of the pump $760-790$~nm, another strong resonance with a Q factor of $52$ is present, represented by a peak in the scattering efficiency spectrum [Fig.~\ref{fig_concept}(c)]. This is dominated by the electric dipole moment of the antenna, although it also contains higher-order multipolar contributions (not shown). The strong internal fields at the Mie-type resonances allow for strong enhancement of the nonlinear frequency mixing processes and also imposes a spectral selection for the frequencies of the generated photons.

The SPDC process in the nanocylinder can result in the emission of photon pairs with nontrivial correlations, associated with different angular and polarization components. In order to experimentally determine the optimal conditions for photon-pair generation and ultimately for optimum SPDC efficiency, usually one uses a technique called quantum state tomography~\cite{James:2001-52312:PRA, Altepeter:2005-105:AAMOP, Lvovsky:2009-299:RMP, Titchener:2018-19:NPJQI}. However, due to a weak efficiency of the spontaneous processes in the small volume of the nanoantenna, the bi-photon rate tends to be low. Thereby long acquisition times of the photon counting statistics are required to obtain sufficient statistical data for significant correlation precision. Therefore, optimizing the experimental parameters directly through SPDC measurements is impractical and an alternative solution is needed.
\subsection{Nonlinear classical characterisation}
\begin{figure}[tb!]
\begin{center}
\includegraphics[width=\textwidth]{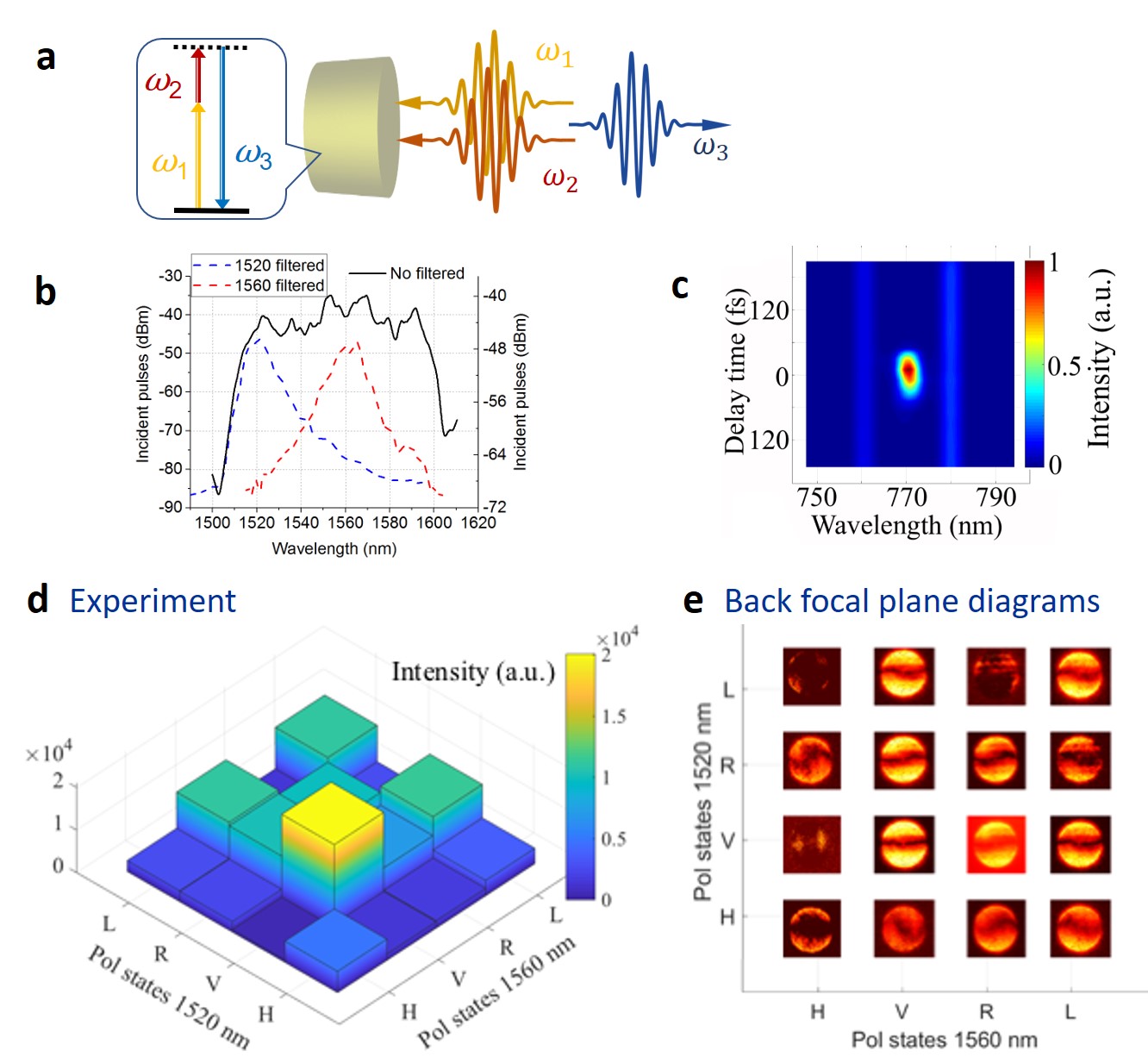}
\caption{SFG nonlinear characterisation of correlations in the nanoantenna. (a) Schematic of the experimental arrangement and energy conservation diagram of the SFG process in the inset. 
(b)~Spectra of the signal (blue-dashed line) and idler (red-dashed line) spectra filtered from the fs laser source. 
(c)~Spectrum of the nonlinear wave mixing in the AlGaAs nanocylinder as a function of the time delay between the signal and idler pulses. The SFG signal is only visible when the two pulses overlap onto the nanoantenna. The spectral features at 760 and 780~nm correspond to the second harmonic generation from the individual signal and idler pulses. 
(d)~Intensity of H-polarized reflected SFG at 770~nm, measured with 16 combinations of horizontal (H), vertical  (V), right circular (R) and left circular (L) polarizations of signal and idler beams for the nanocylinder geometry in Fig.~\ref{fig_concept}. 
(e)~Measured reflected SFG images in $k$-space for the polarization combinations shown in (d) and SFG detected with $NA=0.7$. After Ref.~\cite{Marino:2019-1416:OPT}.}
\label{fig3}
\end{center}
\end{figure}

We employ the concept of quantum-classical correspondence between SPDC and its reversed process, namely SFG, as described in Sec.~\ref{sec:quantum-classical}. We recall that the generated sum-frequency and pump waves should propagate in opposite directions to the SPDC pump, signal and idler~\cite{Poddubny:2016-123901:PRL, Lenzini:2018-17143:LSA}. The quantum-classical correspondence will allow for the optimisation of the excitation parameters, as well as for classical estimation of the SPDC generation bi-photon rates. We consider the collection of photons in all directions and accordingly perform the integration of Eq.~\eqref{eq:corr2} over solid angles within the half-sphere, resulting in the relation:
\begin{equation} \label{eq:1}
   \frac{1}{\Phi_{p}} \frac{d N_{\rm pair}}{dt} =
2\pi\Xi^{\rm SFG} \frac{\lambda_p^4}{\lambda_s^{3}\lambda_i^{3}}
                  \frac{c \Delta\lambda_s}{\lambda_s^{2}} .
\end{equation}
Here, $\Phi_p$ is the SPDC pump flux, $\lambda_p$, $\lambda_s$ and $\lambda_i$ are the pump, signal and idler wavelengths, and $\Delta\lambda_s$ is the nonlinear resonance bandwidth at the signal wavelength. The efficiency $\Xi^{\rm SFG}$ is given by the ratio of the sum-frequency photon power to the product of incident energy fluxes at signal and idler frequencies. 
The number of photon pairs generated through SPDC, in a given optical mode of the nanostructure, is therefore proportional to the SFG amplitude of the classical signal and idler waves, propagating in the opposite direction. In this framework, one can first optimize the SFG efficiency and thus predict the bi-photon generation rates, prior to SPDC detection. Importantly, the SFG process can also be characterized for different polarizations, thereby optimizing the parameters for the subsequent SPDC measurements. 

The schematic of such SFG experiments is illustrated in Fig.~\ref{fig3}(a). Two femtosecond laser pulses at wavelengths 1520~nm and 1560~nm illuminate the nanoantenna as signal and idler beams. Their spectra are shown in Fig.~\ref{fig3}(b). The two beams are focused onto a single AlGaAs nanocylinder by a 0.7~NA objective, with an average power of 10~mW. The beam size of the two incident pulses is $2~\mu$m (diameter) diffraction-limited spots, resulting in 7~GW/cm$^2$ peak intensities. The incident linear H polarization is parallel to the AlGaAs nanoresonator's crystallographic axis $(100)$. Figure~\ref{fig3}(c) shows the time resolved spectra of H-polarized emission collected in backward direction at different optical delays between the two V-polarized pulses. This polarizations arrangement corresponds to the maximum SFG efficiency, as we discuss below. The two spectral peaks at 760 and 780~nm correspond to the second harmonic generation (SHG) from the individual signal and idler pulses, and are observed at all delay times. The third peak at 770~nm only occurs at ``time zero,'' when the signal and idler pulses arrive at the nanocylinder simultaneously. This SFG pulse has a full-width at half-maximum (FWHM) of 80~fs, in agreement with the duration of the pump pulses.

By setting different combinations of incident polarizations for the signal and idler pulses, including horizontal (H), vertical (V), right circular (R) and left circular (L), one can measure the SFG for H (or V) polarization. The choice for H polarized SFG is arbitrary, as for normally incident signal/idler beams V and H are identical due to the cylindrical symmetry of the disk and the isotropy of the material. The resulting SFG signal intensities (normalized to the maximum value) at 770~nm and the corresponding radiation patterns, recorded via a back focal plane (BFP) imaging system, are shown in Figs.~\ref{fig3}(d) and~(e), respectively. The maximum signal of H-polarized SFG is obtained when both signal and idler are V-polarized. At the microscopic scale, this corresponds to the excitation of signal, idler and SFG modes whose vectorial components constructively overlap, following the symmetry of AlGaAs second-order susceptibility tensor. The highest SFG conversion efficiency from the nanocylinder is measured to be $1.8\times10^{-5}$, which is comparable to the SHG efficiency obtained by different groups~\cite{Gili:2016-15965:OE, Liu:2016-5426:NANL, Camacho-Morales:2016-7191:NANL}. 
As shown in the BFP images, the SFG radiation patterns strongly depend on signal and idler polarization combinations, however the general observation is that the SFG signal is emitted under angle, off-axis to the nanocylinder. This is due to the symmetry of the nonlinear tensor, as previously reported for SHG in Refs.~\cite{Carletti:2016-1500:ACSP, Camacho-Morales:2016-7191:NANL}. 

The experimental results have been compared with finite element simulations under realistic experimental conditions. The simulated SFG intensity is enhanced when the polarizations of both signal and idler beams are VV, or RR or LL polarized. Lower counts are seen for the mixed polarization cases, and for the case HH. This trend matches the experimental results, particularly for the combinations involving H and V polarizations, while the RR and LL cases appears less bright than VV case. Such discrepancy can be attributed to slight non-uniformity of the fabricated structure, which can deviate from the cylindrical to elliptical shape.

Importantly, knowing the SFG efficiency of $1.8\times10^{-5}$ for the VV $\rightarrow$ H process, one can estimate the possible bi-photon rates for detection of SPDC photon pair rates from an AlGaAs nanoantenna. The prediction for the photon-pair generation rate, obtained using Eq.~(\ref{eq:1}), is about 380~Hz at a pump power of 2¬mW. This value is significant and well above the dark count rates for the detectors used in the experiments (estimated at 5~Hz). However, for Eq.~(\ref{eq:1}) to exactly predict the following SPDC experiments, one needs to look for SFG emission mainly directed backwards. 
The SFG emission is maximal in the backward direction when the signal and idler beams are incident at oblique angle. Theoretical predictions show that a high normal SHG can be obtained when the signal and idler illuminate the nanoantenna at $45^\circ$ to the nanocylinder axis. An SFG experiment where the illumination is carried out through a high NA objective, will partially  satisfy this criterion. Therefore, under such experimental condition Eq.~(\ref{eq:1}) is expected to overestimate the detected SPDC count rate.
\begin{figure}[tb!]
\begin{center}
\includegraphics[width=0.8\columnwidth]{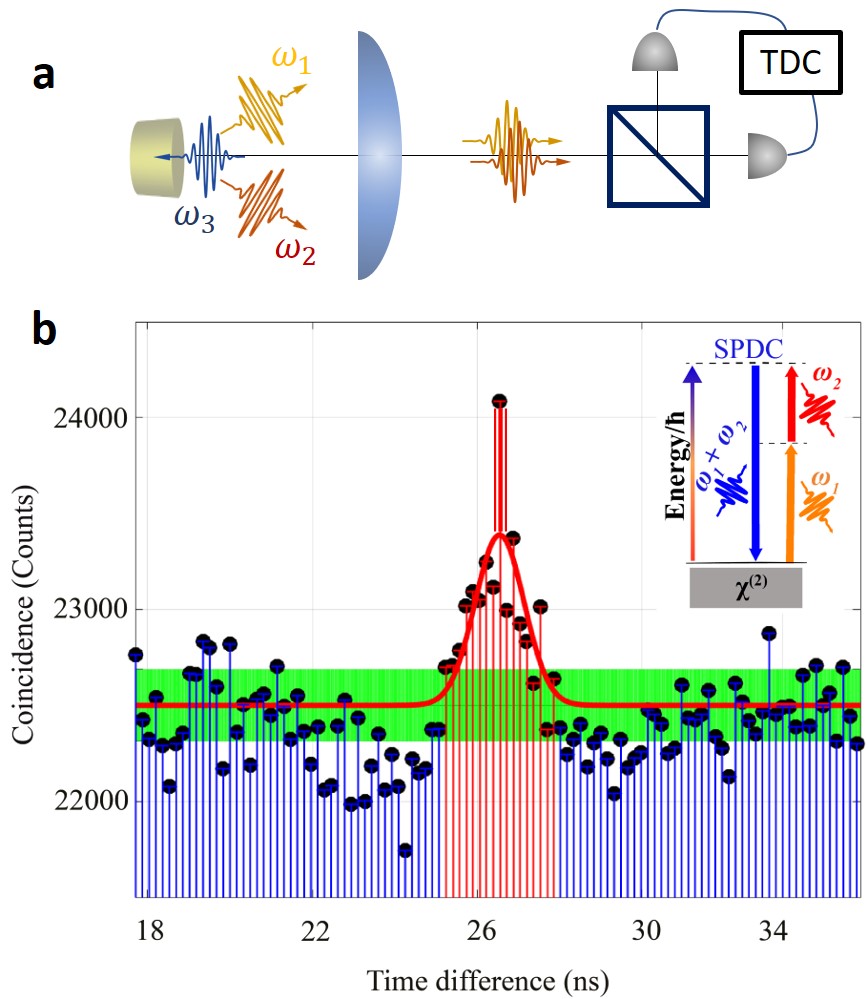}
\caption{(a) Schematics of the experiment for generation of biphoton-pairs in an AlGaAs disk nanoantenna. (b) IR coincidence counts integrated over 24~hours on two single-photon detectors after a beam splitter. A significant statistical increase, marked by the red bar, is apparent at a time difference of 26.5~ns, corresponding to the temporal delay between both detectors. Black dots are the measured coincidences, the red shadowed area indicates correlation due to thermal excitation of the semiconductor materials, while the red solid-line is its fitted Gaussian curve. The inset shows a schematic of the SPDC process and energy correlation. The green band depicts the statistical error of the measurement. After Ref.~\cite{Marino:2019-1416:OPT}.
}
\label{fig4}
\end{center}
\end{figure}

\subsection{SPDC experiments with single nanoantennas}
Once knowing the optimal antenna parameters for enhancement of the SFG process, the generation of photon pairs through the SPDC process can be tested. 
A CW pump laser at 785~nm was incident onto the AlGaAs nanoantenna with a power of 2~mW and a $2~\mu$m (diameter) diffraction-limited spot. The generated photon pairs with frequencies $\omega_1$ and $\omega_2$ are collected by a high-numerical aperture microscope objective and directed through a non-polarising beam splitter at two single photon InGaAs detectors. The coincidence of detection of the two photons is recorded by a time-to-digital electronics (Fig.~\ref{fig4}a). The generated SPDC photons are expected to have a large spectral bandwidth of about 150~nm, due to the broad magnetic dipole resonance in the IR spectral range, as shown in Fig.~\ref{fig_concept}c. This bandwidth is relatively broad with respect to conventional SPDC sources, which typically have sub-nm or few-nm bandwidth. Such broad bandwidth offers a range of advantages, including a short temporal width for timing-critical measurements, such as for temporal entanglement~\cite{Rogers:2016-1754:ACSP}, or for SPDC spectroscopy~\cite{Solntsev:2018-21301:APLP}. It also dictates a sub-100 fs temporal width of the generated photons, which is much shorter than the coincidence time window $\tau_c$.

The measured coincidences for an H-polarized CW pump are presented in Figure~\ref{fig4}b, where photon counting statistics is accumulated by integrating over 24~hours. For a time difference of 26.5~ns, corresponding exactly to the time difference between the signal and idler detection arms, a single bin with high coincidence rate is observed. This is consistent with the physics of SPDC generation of signal and idler photons with the estimated temporal correlations of sub-100~fs. Although it only emerges from the background by a limited number of counts, this peak of coincidence rate is statistically relevant and larger than the statistical error (marked with green band). A second, broader coincidence statistics is also observed underneath the SPDC peak. This Gaussian peak is the indication of correlation due to thermal excitation of the semiconductor materials~\cite{Flagg:2012-163601:PRL}. It has approximately 2~ns width, as shown with the red-shadowed area in Fig.~\ref{fig4}b. The experimental SPDC rate from the AlGaAs nanocylinder has been analysed, taking into account the losses in the detection system. The estimate of the total photon-pair generation rate from the nanoantenna is then estimated to $d N^{gen}_{\rm disk} / dt = 35$~Hz. Normalized to the pump energy stored by the nanoantenna, this rate reaches values of up to 1.4~GHz/Wm, being one order of magnitude higher than conventional on-chip or bulk photon-pair sources~\cite{Kwiat:1995-4337:PRL,Guo:2017-e16249:LSA}.
\begin{figure}[!htb]
\begin{center}
\includegraphics[width=0.9\columnwidth]{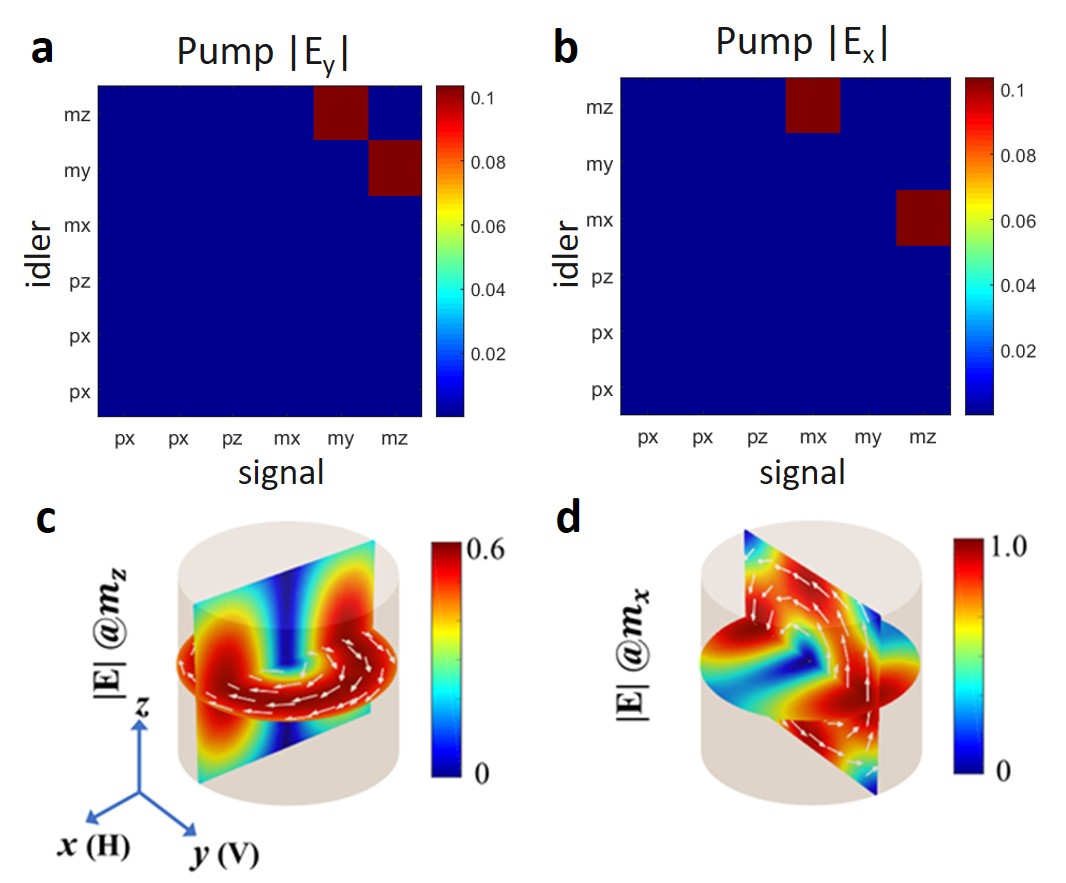}
\vspace*{-1mm}
\caption{(a,b) Mode correlation in the AlGaAs nanocylinder for H (left) and V (right) polarized pump excitation. (c,d) Numerically simulated fields inside the nanocylinder when exciting $m_x$ and $m_z$ modes, respectively. The white arrows indicate the electric field vector. Different color bar scales are used, following the different intensities of $m_x$, $m_z$ inside the nanocylinder, in contrast to the symmetric case of a nanosphere. After Ref.~\cite{Marino:2019-1416:OPT}.}
\label{fig:entanglement}
\end{center}
\end{figure}

Importantly, this rate is significantly higher than the reference measurements of the AlOx/GaAs substrate without the nanocylinder that is of the order of $dN^{gen}_{\rm sub}/dt =9$~Hz. The figure of merit calculated for this substrate source results is four-orders of magnitude smaller than for the nanocylinder. While the AlGaAs nanocylinder is weak in absolute values, such a figure of merit brings to light the nanometer spatial confinement and relatively high Q factors achievable in our AlGaAs nanocylinder. The latter operating in the Mie scattering regime, enables a fine shaping of the spectral and radiation profile, thereby leading to flexible quantum state engineering and possible spatial multiplexing of SPDC sources in a metasurface. Note that the influence of the AlGaAs nanocylinder on the SPDC from the substrate, such as refocusing the pump beam or the photon pairs, was found to be negligible by appropriate numerical modeling~\cite{Marino:2019-1416:OPT}.

Finally, it is possible to numerically calculate the sub-wavelength mode correlation responsible for the generation of signal and idler photon pairs (Fig.~\ref{fig:entanglement}a,b). This can be done again through the quantum-classical analogy presented in Sec~\ref{sec:correspond} by calculating the SFG normal output when exciting the idler and signal fields with combination of ED ($p_x$ or $p_y$ or $p_z$) and MD ($m_x$ or $m_y$ or $m_z$) inside the disk at idler or signal wavelengths, respectively. 
Only orthogonal Cartesian components of magnetic dipole contribute to SPDC process, which leads to coupling of two magnetic dipole moments of the nanoantenna, namely $m_x$ and $m_z$. The two coupled modes for an H-polarised pump beam are shown in Fig.~\ref{fig:entanglement}c and d. The coupling of these two sub-wavelength modes efficiently generates photon pairs in the far-field via the antenna radiation and underpins the measured photon correlation.
As can be seen from Fig.~\ref{fig:entanglement}a,b, with $\bm E_H \parallel y$ ($\bm E_V \parallel x$) pump, the generated two photons are dominantly coupled into $m_x$ and $m_z$ ($m_y$ and $m_z$) modes, respectively. Due to the SFG-SPDC correspondence, the SFG process has the same symmetry properties and can be understood using the group theory~\cite{Dresselhaus:2008:GroupTheory}. At the signal and idler frequencies, the electric field is controlled by the magnetic dipole resonance, while at the pump/SFG frequency it is determined by the electric dipole mode. In the $T_d$ symmetry group of the AlGaAs crystalline lattice, the magnetic dipole modes transform according to the $F_1$ irreducible representation, while the electric dipole modes belong to the $F_2$ representation. The basis functions of the $F_2$ representation, corresponding to the direct product 
$F_1\otimes F_1$ are $p_x\propto m_y m_z,
p_y\propto m_x m_z,p_z\propto m_x m_y$~\cite{Dresselhaus:2008:GroupTheory} in full agreement with the numerical calculations in Fig.~\ref{fig:entanglement}. The states shown in Figs.~\ref{fig:entanglement}(a,b) have the corresponding Schmidt number of 2, indicating a very strong correlation between the modes. More detailed symmetry analysis of the SFG process in dielectric nanoparticles  can be found in Ref.~\cite{Frizyuk:2019-75425:PRB}.

We note that similar SPDC nanoscale sources can be obtained with other nonlinear crystalline nanostructures, including lithium niobate~\cite{Carletti:2019-33391:OE}. This would allow one to explore different crystalline symmetries and polarisation dependencies. As such, the field of nanoscale sources of two-photon quantum states is expected to grow in the years to come. Therefore, the development of nanoscale quantum sources based on SPDC is likely to be of interest to the wider quantum community.
\section{Outlook}\label{sec:outlook}
\begin{figure}[tb]
\begin{center}
\includegraphics[width=0.9\columnwidth]{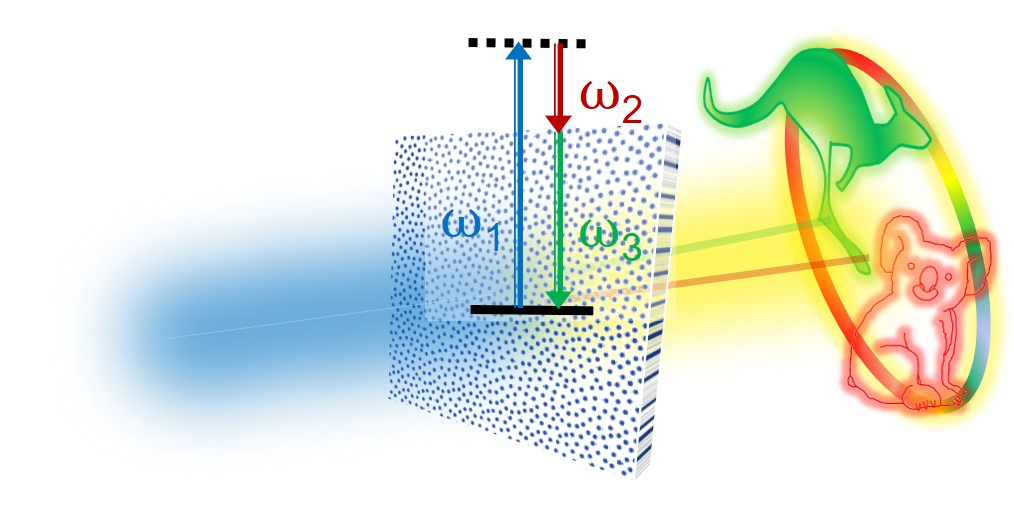}
\vspace*{-1mm}
\caption{Vision for a photon-pair generation via down-conversion process in a spatially-variant nonlinear metasurface, enabling the generation of correlated images.}
\label{fig:outlook}
\end{center}
\end{figure}

The experiments on SPDC in a single AlGaAs nanodisk bring confidence that correlated photons can be generated by employing crystalline nanoantennas and arrangement of such nanoantennas in an optical metasurface. Such nanoscale platform can open new opportunities for generation, unrestricted by longitudinal phase matching~\cite{Okoth:2019-263602:PRL}, of non-classical spatially entangled states of arbitrary shapes by carefully engineering of the dimensions of different nanoantennas in spatially variant metasurfaces, see artistic image in Fig.~\ref{fig:outlook}. 
We believe that this opportunity, combined with the capacity of metasurfaces to transform, image and reconstruct quantum states~\cite{Stav:2018-1101:SCI, Wang:2018-1104:SCI, Altuzarra:2019-20101:PRA, Georgi:2019-70:LSA}, will unleash a potential for ultimate miniaturization of quantum devices suitable for end-user applications~\cite{Silverstone:2014-104:NPHOT, Xiong:2016-10853:NCOM}, such us quantum imaging~\cite{Lemos:2014-409:NAT}, sensing~\cite{Saravi:2017-4724:OL}, precision spectroscopy~\cite{Kalashnikov:2016-98:NPHOT}, free-space communications~\cite{Willner:2017-20150439:PTRSA} and cryptography~\cite{Xiong:2016-10853:NCOM}.

The authors acknowledge highly productive collaborations, and in particular the results outlined in this chapter were primarily driven by researchers who co-authored Refs.~\cite{Poddubny:2016-123901:PRL, Lenzini:2018-17143:LSA, Marino:2019-1416:OPT}.
\providecommand{\WileyBibTextsc}{}
\let\textsc\WileyBibTextsc
\providecommand{\othercit}{}
\providecommand{\jr}[1]{#1}
\providecommand{\etal}{~et~al.}

\end{document}